\documentclass{elsart}%
\usepackage{amsmath}
\usepackage{amssymb}
\usepackage{graphicx,amssymb}
\usepackage[margin=10pt,font=small,labelfont=bf,labelsep=endash]{caption}
\usepackage{amsfonts}%
\usepackage{graphicx}
\providecommand{\U}[1]{\protect\rule{.1in}{.1in}}
\providecommand{\U}[1]{\protect\rule{.1in}{.1in}}
\begin{document}
\begin{frontmatter}
\title{Numerical Methods for a Poisson-Nernst-Planck-Fermi Model}
\author{Jinn-Liang Liu}
\address{Department of Applied Mathematics,
National Hsinchu University of Education, Hsinchu 300, Taiwan
(jinnliu@mail.nhcue.edu.tw)}
\author{Bob Eisenberg}
\address{Department of Molecular Biophysics and Physiology,
Rush University, Chicago, IL 60612 USA (beisenbe@rush.edu)}
\end{frontmatter}

\textbf{Abstract.} Numerical methods are proposed for an advanced
Poisson-Nernst-Planck-Fermi (PNPF) model for studying ion transport through
biological ion channels. PNPF contains many more correlations than most models
and simulations of channels, because it includes water and calculates
dielectric properties consistently as outputs. This model accounts for the
steric effect of ions and water molecules with different sizes and
interstitial voids, the correlation effect of crowded ions with different
valences, and the screening effect of polarized water molecules in an
inhomogeneous aqueous electrolyte. The steric energy is shown to be comparable
to the electrical energy under physiological conditions, demonstrating the
crucial role of the excluded volume of particles and the voids in the natural
function of channel proteins. Water is shown to play a critical role in both
correlation and steric effects in the model. We extend the classical
Scharfetter-Gummel (SG) method for semiconductor devices to include the steric
potential for ion channels, which is a fundamental physical property not
present in semiconductors. Together with a simplified matched interface and
boundary (SMIB) method for treating molecular surfaces and singular charges of
channel proteins, the extended SG method is shown to exhibit important
features in flow simulations such as optimal convergence, efficient nonlinear
iterations, and physical conservation. The generalized SG stability condition
shows why the standard discretization (without SG exponential fitting) of NP
equations may fail and that divalent Ca$^{2+}$ may cause more unstable
discrete Ca$^{2+}$ fluxes than that of monovalent Na$^{+}$. Two different
methods --- called the SMIB and multiscale methods --- are proposed for two
different types of channels, namely, the gramicidin A channel and an L-type
calcium channel, depending on whether water is allowed to pass through the
channel. Numerical methods are first validated with constructed models whose
exact solutions are known. The experimental data of both channels are then
used to verify and explain novel features of PNPF as compared with previous
PNP models. The PNPF currents are in accord with the experimental I-V (V for
applied voltages) data of the gramicidin A channel and I-C (C for bath
concentrations) data of the calcium channel with 10$^{-8}$-fold bath
concentrations that pose severe challenges in theoretical simulations.\bigskip

PACS number(s): 87.10.Ed 02.70.-c 47.61.Cb 05.20.Jj\newline

\begin{center}
\textbf{I. INTRODUCTION}\\[0pt]

\bigskip
\end{center}

The literature on numerical methods for drift-diffusion (DD) or
Poisson-Nernst-Planck (PNP) models of semiconductor devices and ion channels
is large, including
\cite{G64,SG69,S73,BR83,MR83,S84,S88,K88,AM89,BM89,RS89,AR93,J95,KC99,CC00,CL03,P04,CL05,CL08,LH10,SK10,ZC11}
and references therein. In biological simulations, continuum models have been
challenged as inaccurate compared to Monte Carlo (MC), Brownian dynamics (BD),
or molecular dynamics (MD) due to the gross approximation of atomic properties
of channel proteins and electrolyte solutions
\cite{SD99,AK00,CK00,SW01,IR01,IR02,CK02,EC02,MJ04,RA04,SC11}. Continuum
models on the other hand have substantial advantages in efficiency that are of
great importance in studying a range of conditions and concentrations
especially for large nonequilibrium or inhomogeneous systems, as are present
in experiments and in life itself
\cite{CC00,LH10,SK10,ZC11,IR02,FB02,GK04,CK05,E10,E11,E12,WZ12,E13}.

Based on the configurational entropy model \cite{LE13} for aqueous
electrolytes with arbitrary $K$ species of nonuniform size, hard spherical
ions, we extended the Poisson-Fermi model in \cite{LE13} to a new model ---
called the Poisson-Nernst-Planck-Fermi (PNPF) model --- for nonequilibrium
systems by including specifically the excluded volume effects of the next
species ($K+1$) of water molecules and the interstitial voids ($K+2$) between
all particles \cite{LE14b}. The PNPF model differs from most channel models in
several respects: (i) it computes dielectric properties as an output that in
fact vary with position and with experimental condition, (ii) a fourth order
Cahn-Hilliard type partial differential equation emerges to replace the second
order Poisson equation of PNP, which has a richness of behavior beyond the
usual second order PNP description, and (iii) using the methods of this paper,
this more powerfully correlated model is in fact much easier to compute in
three dimensions than other steric PNP models. Previous work \cite{LE14b}
gives more details.

The PNPF model also provides a quantitative mean-field description of the
charge/space competition mechanism of particles within the highly charged and
crowded channel pore. The steric energy lumps the effects of excluded volumes
of all ions, water, and voids. It yields an energy landscape of ions that
varies significantly with bath concentrations in a 10$^{8}$-fold range of
experimental conditions for L-type calcium channels.

\textbf{\underline{\textbf{Computational Challenges.}}} The 10$^{8}$-fold
range of bath conditions and the highly energetic behavior of permeating ions
through the extremely crowded narrow channels pose severe challenges in
implementations. The strength of local electric fields in a calcium channel
can be higher than that in a semiconductor device (comparing, for example,
$0.27$ V/nm estimated from Fig. 13 in this paper for ion channel and $0.06$
V/nm from Fig. 7 in \cite{CL05} for semiconductor device). This means that the
convergence and stability problems in ion channel simulation can be more
severe than those in semiconductor devices. These problems are not made easier
by the presence of countervailing steric potentials of the same order of magnitude.

Moreover, geometric complexity and singularities of molecular surfaces
separating electrolyte solutions from protein atoms in biological systems need
to be carefully treated in order to obtain tolerable accuracy in 3D PNP
simulations \cite{ZC11}. Seemingly small numerical over approximations can
lead to errors that make results not useful. A second-order method called the
matched interface and boundary (MIB) method was developed by Wei et al.
\cite{ZC11,GY07} for Poisson-Boltzmann and PNP models and is simplified (SMIB)
in \cite{L13} for the PF model to deal with the geometric singularities by the
standard finite difference approximation.

The Scharfetter-Gummel (SG) \cite{SG69} method is an \textit{optimal} and
uniformly convergent method (with respect to the mesh size) to discretize
drift-diffusion (or Nernst-Planck) equations for flux calculations because it
integrates the corresponding 1D initial value problem \textit{exactly} at
\textit{every} grid point \cite{MR83}. We extend the classical SG method to
the NPF equation by showing how the Fermi distribution of hard spheres of
water and ions is imposed. If the classical Boltzmann distribution is used,
the density of point charges would grossly overestimate ionic concentrations
(that are in fact limited because of the finite size of ions) and consequently
lead to inaccurate electrostatic potential and ion mobility by the classical
PNP \cite{CK00,CK02,GK04}. We also show that the classical
Goldman-Hodgkin-Katz flux approximation \cite{H01} in ion channels is in fact
exactly the Scharfetter-Gummel flux approximation on grid points in
semiconductor devices. Similar results appear in the seminal work of Mott
\cite{M39} that was well known to Hodgkin, Cole, and Goldman. The pioneers in
two different fields had the same idea that made a profound impact on their
respective fields and others.

The SG stability condition --- a critical condition of the flux equation in
implementation --- is also extended to include the steric potential that is
not present in classical PNP models. This stability condition explains why the
standard finite difference or finite element discretization fails when the
electric and/or steric potentials vary sharply in a layer region and the mesh
of grid points is not sufficiently resolved. It plays a key role in preserving
physically positive concentrations and divergence-free currents (current
conservation) in approximation \cite{BM89}. We take a closer look at the
numerics concerning the extended SG condition and discover that this condition
is harder to satisfy for the standard methods for the divalent Ca$^{2+}$ than
the monovalent Na$^{+}$ flux since the SG condition depends on the valence of
ions. This is physically reasonable because Ca$^{2+}$ ions are more energetic
in binding and permeation in voltage-gated calcium channels that conduct
Ca$^{2+}$ ions with high-fidelity and high-throughput \cite{SM03}.

The combined method --- the SMIB-SG method --- is shown not only to achieve
second order of accuracy for the PNPF model (with constructed exact solutions)
but also to outperform the primitive SMIB method (without SG) for the
gramicidin A channel due to the exactness property of the SG exponential
fitting between grid points. We also show that the primitive SMIB method fails
to converge for the calcium channel due to its highly charged (-4$e$, $e$ is
the proton charge) and very narrow (about 1 \AA \ in radius) binding site as
compared with that of the gramicidin A channel (-2$e$ and about 2 \AA \ in
radius). In our simulations, water (1.4 \AA \ in radius) is found to flow
through the gramicidin A channel but not to flow through the calcium channel
in some conditions. We use a second method --- called the multiscale method
--- that treats water and ions explicitly in the binding site of the calcium
channel so that water may not move through the channel. It is multiscale since
both Poisson's theory of continuous charges and Coulomb's law of discrete
charges are used in the solvent domain. This demonstrates the novelty of the
PNPF model as compared with previous PNP models in dealing with ion-protein,
ion-ion, and ion-water interactions and the steric effect of ions and water in
the narrow pore. PNPF captures many more of the correlations not present in
PNP itself. It captures steric interactions of ions and water and packs them
well (i.e. consistently) because it includes free space. Dielectric properties
vary with position and concentration and are fully consistent with the rest of
the model because they are outputs of the calcualtions, not inputs, as assumed
in most channel models.

The nonlinear algebraic systems of discrete PNP equations are very difficult
to solve due to strong nonlinearity of the coupled system in both
semiconductor devices and ion channels, especially with sharp potentials at
practical applied voltages
\cite{MR83,AM89,BM89,KC99,CC00,CL03,CL05,CL08,LH10,SK10,ZC11}. The PNPF model
consists of $K+2$ PDEs (1 fourth-order Poisson-Fermi and $K+1$ second-order
Nernst-Planck). The fourth-order PF equation was proposed to account for the
correlation effect of ions in water \cite{S06} and transformed to two
second-order PDEs for computational efficiency and for calculating variable
permittivity within the channel pore \cite{LE13,L13}. The last NP equation
describes the dynamics of water molecules that play a critical role not only
in the steric arrangement of all particles but also in its screening and
polarization effects on ions in the system \cite{LE13,L13}. The full PNPF
model incorporates these atomic properties and thus can provide more accurate
simulations but obviously at the expense of more difficulties in
implementation than that of previous PNP models. It is impractical to solve
the $(K+2)M$ nonlinear algebraic equations resulting from a discretization of
PNPF using Newton's iteration on the coupled system, where the matrix size $M$
corresponding to each PDE can easily grow to millions in 3D implementations.
With a linearized Poisson equation, Gummel's iteration is an efficient method
because it solves each PDE successively \cite{AM89}. It has been shown in
\cite{ZC11} that an SOR-like method (without linearization) converges faster
than Gummel's method at higher bath concentrations for ion channel simulations
provided that the relaxation parameter is appropriately chosen. We present a
new SOR-like method for the PNPF model that differs from the previous models
in the fourth-order PDE, the water NP equation, and the steric potential. It
is shown that the method improves the convergence rate using the same
gramicidin A channel protein as considered in \cite{ZC11}.

The rest of the paper is organized as follows. In Section 2, we briefly
describe the PNPF theory. Sections 3 and 4 present the numerical methods
proposed in this paper. Section 4 also include two algorithms with respect to
the SMIB and multiscale methods to illustrate implementation procedures for
studying two different types of ion channels. In Section 5, the SMIB and SG
methods are first validated by using a real protein structure of the GA
channel with a set of exact solutions constructed for the PNP model. Both
methods are shown to achieve optimal results as analyzed in this paper. The
extended SG condition is then carefully scrutinized in discretization and used
to explain why the standard discretization method is not feasible for the
calcium channel model considered here, especially to approximate the high
energetic Ca$^{2+}$ flux. PNPF results are shown to agree with experimental
I-V and I-C data of GA and calcium channels using the two algorithms,
respectively. Some concluding remarks are made in Section 6.

\begin{center}
\textbf{II. POISSON-NERNST-PLANCK-FERMI MODEL}
\end{center}

For an electrolyte in a solvent domain $\Omega_{s}$ with arbitrary $K$ species
of ions and the next species $K+1$ of water, the configurational entropy model
proposed in \cite{LE13} is extended in \cite{LE14b} to treat all particles as
hard spheres with nonuniform sizes and to include explicitly as its last
species $K+2$ the voids between all particles. Based on the extended entropy
model, the following Gibbs-Fermi free energy functional of the system is
proposed in \cite{LE14b}
\begin{align}
G^{\text{Fermi}}  &  =\int_{\Omega_{s}}d\mathbf{r}\left\{  -\frac{\epsilon
_{s}l_{c}^{2}}{2}\left(  \nabla^{2}\phi(\mathbf{r})\right)  ^{2}%
-\frac{\epsilon_{s}}{2}\left\vert \nabla\phi(\mathbf{r})\right\vert ^{2}%
+\rho(\mathbf{r})\phi(\mathbf{r})+g\right\} \tag{1}\\
g  &  =k_{B}T\left(  \sum_{j=1}^{K+1}\left[
\begin{array}
[c]{l}%
C_{j}(\mathbf{r})\ln\left(  v_{j}C_{j}(\mathbf{r})\right)  -C_{j}%
(\mathbf{r})\\
-C_{j}(\mathbf{r})\ln\left(  v_{K+2}C_{K+2}(\mathbf{r})\right)  -\frac{\mu
_{i}^{\text{B}}C_{j}(\mathbf{r})}{k_{B}T}%
\end{array}
\right]  \right)  ,\nonumber
\end{align}
where $\epsilon_{s}=\epsilon_{\text{w}}\epsilon_{0}$, $\epsilon_{\text{w}}$ is
the dielectric constant of bulk water, $\epsilon_{0}$ is the vacuum
permittivity, $l_{c}$ is a correlation length \cite{S06,BS11}, $\phi
(\mathbf{r})$ is the electrostatic potential function of spatial variable
$\mathbf{r}\in$ $\Omega_{s}$, $\rho(\mathbf{r})=\sum_{j=1}^{K+1}q_{j}%
C_{j}(\mathbf{r})$ is the charge density, $C_{j}(\mathbf{r})$ is the
concentration of type $j$ particles carrying the charge $q_{j}$ $=z_{j}e$ with
valence $z_{j}$ and having the volume $v_{j}=4\pi a_{j}^{3}/3$ with radius
$a_{j}$, $k_{B}$ is the Boltzmann constant, $T$ is the absolute temperature,
and $\mu_{i}^{\text{B}}=k_{B}T\ln\left(  v_{i}C_{i}^{\text{B}}/\Gamma
^{\text{B}}\right)  $ is a constant chemical potential. Water is treated as
polarizable spheres with zero net charge, so $z_{K+1}=q_{K+1}=0$.

The total volume $V$ of the system consists of the volumes of all particles
and the total void volume $v_{K+2}$, i.e., $V=\sum_{j=1}^{K+1}v_{j}%
N_{j}+v_{K+2}$, where $N_{j}$ is the total number of type $j$ particles. Under
the bulk condition, dividing this equation by $V$ yields the bath void volume
fraction
\begin{equation}
\Gamma^{\text{B}}=\frac{v_{K+2}}{V}=1-\sum_{j=1}^{K+1}v_{j}\frac{N_{j}}%
{V}=1-\sum_{j=1}^{K+1}v_{j}C_{j}^{\text{B}}\text{,}\tag{2}%
\end{equation}
where $C_{j}^{\text{B}}$ is the bath concentration. The void fraction function%
\begin{equation}
\Gamma(\mathbf{r)}=1-\sum_{j=1}^{K+1}v_{j}C_{j}(\mathbf{r})=v_{K+2}%
C_{K+2}(\mathbf{r})\text{,}\tag{3}%
\end{equation}
varies with concentrations $C_{j}(\mathbf{r})$ of all particles and thus with
the distribution $C_{K+2}(\mathbf{r})$ of interstitial voids.

Minimizing the Gibbs-Fermi functional (1) with respect to $\phi$ and $C_{i}$
yields the Poisson-Fermi equation \cite{LE13,L13,S06,BS11}
\begin{equation}
\epsilon_{s}\left(  l_{c}^{2}\nabla^{2}-1\right)  \nabla^{2}\phi
(\mathbf{r})=\sum_{i=1}^{K}q_{i}C_{i}(\mathbf{r})=\rho(\mathbf{r})\tag{4}%
\end{equation}
and the Fermi distribution%
\begin{equation}
C_{i}(\mathbf{r})=C_{i}^{\text{B}}\exp\left(  -\beta_{i}\phi(\mathbf{r}%
)+S^{\text{trc}}(\mathbf{r})\right)  \text{, \ \ }S^{\text{trc}}%
(\mathbf{r})=\ln\frac{\Gamma(\mathbf{r)}}{\Gamma^{\text{B}}}\text{,}\tag{5}%
\end{equation}
respectively, where $\beta_{i}=q_{i}/(k_{B}T)$ and $S^{\text{trc}}%
(\mathbf{r})$ is called the steric potential. The fourth-order PF equation
reduces to the classical Poisson-Boltzmann (PB) equation $-\epsilon_{s}%
\nabla^{2}\phi=\rho$ and the Fermi distribution reduces to the Boltzmann
distribution $C_{i}=C_{i}^{\text{B}}\exp\left(  -\beta_{i}\phi\right)  $ when
$l_{c}=S^{\text{trc}}(\mathbf{r})=0$. The distribution (5) is of Fermi type
since all concentration functions are bounded above, $C_{i}(\mathbf{r}%
)<1/v_{i}$ \cite{LE14b}, i.e., $C_{i}(\mathbf{r})$ cannot exceed the maximum
value $1/v_{i}$ for any arbitrary (or even infinite) potential $\phi
(\mathbf{r})$ at any location $\mathbf{r}$ in the domain $\Omega_{s}$.

If $l_{c}\neq0$, the dielectric operator $\widehat{\epsilon}=\epsilon
_{s}(1-l_{c}^{2}\nabla^{2})$ approximates the permittivity of the bulk solvent
and the linear response of correlated ions \cite{BS11}. The dielectric
function $\widetilde{\epsilon}(\mathbf{r})=\epsilon_{\text{w}}/(1+\eta/\rho)$
is a further approximation of $\widehat{\epsilon}$. It is found by
transforming (4) into two second-order PDEs \cite{L13}%
\begin{align}
\text{PF1 }  &  \text{:}\text{ }\epsilon_{s}\left(  l_{c}^{2}\nabla
^{2}-1\right)  \Psi(\mathbf{r})=\rho(\mathbf{r})\tag{6}\\
\text{PF2 }  &  \text{:}\text{\ \ \ \ \ \ \ \ \ \ \ \ \ \ \ \ }\nabla^{2}%
\phi(\mathbf{r})=\Psi(\mathbf{r})\tag{7}%
\end{align}
by introducing a density like variable $\Psi$ that yields a polarization
charge density $\eta=-\epsilon_{s}\Psi-\rho$ of water using Maxwell's first
equation \cite{LE13}. Numerical approximation of the fourth order equation (4)
was simplified to the standard 7-point finite difference approximation of the
second order equations (6) and (7) in \cite{L13}. Boundary conditions of the
new variable $\Psi$ on the solvent boundary $\partial\Omega_{s}$ were derived
from the global charge neutrality condition \cite{L13}. These functions make
dielectric properties outputs in our model and calculations, unlike in most
other treatments of channels.

Including the electrostatic effect of a total of $Q$ fixed atomic charges
$q_{j}$ located at $\mathbf{r}_{j}$ in the biomolecular domain $\Omega_{m}$
that contains both channel protein and membrane lipids, the PF equation (4) is
written as%

\begin{equation}
\epsilon\left(  l_{c}^{2}\nabla^{2}-1\right)  \nabla^{2}\phi(\mathbf{r}%
)=\sum_{j=1}^{Q}q_{j}\delta(\mathbf{r}-\mathbf{r}_{j})+\sum_{i=1}^{K}%
q_{i}C_{i}(\mathbf{r})=\rho(\mathbf{r})\text{, }\forall\mathbf{r}\in
\Omega\text{,}\tag{8}%
\end{equation}
where $\Omega=\Omega_{s}\cup\Omega_{m}$ and $\delta(\mathbf{r}-\mathbf{r}%
_{j})$ is the delta function. Note that $\epsilon=\epsilon_{m}\epsilon_{0}$,
$l_{c}=0$, $\rho(\mathbf{r})=\sum_{j=1}^{Q}q_{j}\delta(\mathbf{r}%
-\mathbf{r}_{j})$ in $\Omega_{m}$ and $\epsilon=\epsilon_{s}\epsilon_{0}$,
$l_{c}\neq0$, $\rho(\mathbf{r})=\sum_{i=1}^{K}q_{i}C_{i}(\mathbf{r})$ in
$\Omega_{s}$, where $\epsilon_{m}$ is the dielectric constant of biomolecules.
As mentioned above, numerical implementation of Eq. (8) (or Eqs. (6) and (7))
is complicated by the complex molecular surface $\partial\Omega_{m}$ in real
protein structures on which suitable interface conditions for the unknown
functions $\Psi(\mathbf{r})$ and $\phi(\mathbf{r})$ should be properly imposed
\cite{L13}. The approximation of interface conditions is not straightforward
\cite{ZC11,GY07,L13} and can be made much worse by geometric singularities of
$\partial\Omega_{m}$ if the singularities are not properly treated. It was
shown in \cite{HL05} that the standard, second-order finite difference method
is degraded to only $O(h^{0.37})$ by this kind of singularities, where $h$ is
the mesh size of grid points.

For nonequilibrium systems, the classical Poisson-Nernst-Planck model
\cite{CB92,EC93,EK95} can then be generalized to the
Poisson-Nernst-Planck-Fermi model by coupling the flux density equation (in
steady state)
\begin{equation}
-\nabla\cdot\mathbf{J}_{i}(\mathbf{r})=0,\text{ }\mathbf{r}\in\Omega
_{s}\tag{9}%
\end{equation}
of each particle species $i=1,\cdots,K+1$ (including water) to the PF equation
(8), where the flux density is defined as%
\begin{equation}
\mathbf{J}_{i}(\mathbf{r})=-D_{i}\left[  \nabla C_{i}(\mathbf{r})+\beta
_{i}C_{i}(\mathbf{r})\nabla\phi(\mathbf{r})-C_{i}(\mathbf{r})\nabla
S^{\text{trc}}(\mathbf{r})\right] \tag{10}%
\end{equation}
and $D_{i}$ is the diffusion coefficient. The flux equation (9) is called the
Nernst-Planck-Fermi equation because the Fermi steric potential $S^{\text{trc}%
}(\mathbf{r})$ is introduced to the classical NP equation. The NPF equation
(9) reduces to the Fermi distribution (5) at equilibrium \cite{LE14b}.

The gradient of the steric potential $\nabla S^{\text{trc}}$ in (10)
represents an entropic force of vacancies exerted on particles. The negative
sign in $-C_{i}\nabla S^{\text{trc}}$ means that the steric force $\nabla
S^{\text{trc}}$ is in the opposite direction to the `diffusional' force
$\nabla C_{i}$, i.e., the larger $S^{\text{trc}}=\ln\frac{\Gamma(\mathbf{r)}%
}{\Gamma^{\text{B}}}$ (meaning more space available to the particle as implied
by the numerator) at $\mathbf{r}$ in comparison with that of neighboring
locations, the more the entropic force pushes the particle to the location
$\mathbf{r}$. The entropic force is simply opposite to the diffusional force
$\nabla C_{i}$ that pushes the particle away from $\mathbf{r}$ if the
concentration is larger at $\mathbf{r}$ than that of neighboring locations.
Moreover, the Nernst-Einstein relationship \cite{H01} implies that the steric
flux $D_{i}C_{i}\nabla S^{\text{trc}}$ is greater if the particle is more
mobile. Therefore, the gradients of electric and steric potentials
($\nabla\phi$ and $\nabla S^{\text{trc}}$) describe the charge/space
competition mechanism of particles in a crowded region within a mean-field
framework \cite{LE14b}. For more physical and mathematical details about the
PNPF theory, we refer to \cite{LE14b}.

\begin{center}
\textbf{III. A GENERALIZED SCHARFETTER-GUMMEL METHOD\bigskip}
\end{center}

We use the standard 7-point finite difference (FD) scheme in 3D \cite{L13} to
discretize the PNPF model. For ease of notation, we omit the subscript $i$ in
(9) when no confusion should arise. For conciseness, the FD discretization is
simplified to 1D in the following discussions as the corresponding 3D case
follows obviously in a similar way. Furthermore, we only provide the FD
formula for the flux equation (9) as the FD formulas with the SMIB method
across the molecular surface $\partial\Omega_{m}$ for Eqs. (6) and (7) have
been given in \cite{L13}, i.e., we consider%
\begin{equation}
\frac{dJ(x)}{dx}=\frac{d}{dx}\left[  -D(x)\left(  \frac{dC(x)}{dx}+\beta
C(x)\frac{d\phi(x)}{dx}-C(x)\frac{dS^{\text{trc}}(x)}{dx}\right)  \right]
=0.\tag{11}%
\end{equation}
The primitive FD approximation of (11) is%
\begin{equation}
\frac{a_{i-1}C_{i-1}+a_{i}C_{i}+a_{i+1}C_{i+1}}{\Delta x^{2}}=0\text{,}%
\tag{12}%
\end{equation}
where $\Delta x=x_{i+1}-x_{i}=h$ is the mesh size of a uniform grid on the $x
$-axis in the domain, $C_{i}\approx C(x_{i})$ is the unknown approximation of
the concentration function $C(x)$ at any grid point $x_{i}$, and the
coefficients are given as%
\begin{equation}
\left\{
\begin{array}
[c]{l}%
a_{i-1}=D_{i-\frac{1}{2}}\left[  1-\beta\Delta\phi_{i-1}/2+\Delta
S_{i-1}^{\text{trc}}/2\right] \\
\text{ \ \ }a_{i}=a_{i-1}+a_{i+1}-2(D_{i-\frac{1}{2}}+D_{i+\frac{1}{2}})\\
a_{i+1}=D_{i+\frac{1}{2}}\left[  1+\beta\Delta\phi_{i}/2-\Delta S_{i}%
^{\text{trc}}/2\right]  ,
\end{array}
\right. \tag{13}%
\end{equation}
where $\Delta\phi_{i-1}=\phi_{i}-\phi_{i-1}$, $\phi_{i}\approx\phi(x_{i})$,
$x_{i+\frac{1}{2}}=(x_{i+1}+x_{i})/2$ etc. The diffusion coefficient function
$D(x)$ is equal to a constant $D^{\text{B}}$ in the bath and to a reduced
constant $\theta D^{\text{B}}$ in the channel pore with $0<\theta<1$. The
function $D(x)$ along the channel axis is constructed by using the
interpolation method presented in \cite{ZC11} for connecting the bath value
$D^{\text{B}}$ and the pore value $\theta D^{\text{B}}$ such that $D(x)$ is a
continuously differentiable function. The factor $\theta$ is the only tuning
parameter in the PNPF model to fit experimental data
\cite{CC00,ZC11,IR02,GK04,CK05,WZ12,FZ06,G08}. We shall investigate the
magnitude of $\theta$ for GA channel and compare it with those obtained by MD
and BD simulations. The comparison is used to verify the correlation and
steric effects considered in PNPF.

At any two adjacent grid points $x_{i}$ and $x_{i+1}$, the FD approximation of
the zero flux ($J(x)=0$) is%
\begin{equation}
\frac{C_{i+1}-C_{i}}{\Delta x}=\frac{C_{i+1}+C_{i}}{2}\left(  -\beta
\frac{\Delta\phi_{i}}{\Delta x}+\frac{\Delta S_{i}^{\text{trc}}}{\Delta
x}\right)  ,\tag{14}%
\end{equation}
which implies that we may obtain the inequality%
\begin{equation}
C_{i+1}-C_{i}>C_{i+1}+C_{i}\tag{15}%
\end{equation}
and thereby a negative (\textit{unphysical}) concentration $C_{i}<0$ at
$x_{i}$ if%
\begin{equation}
\frac{1}{2}\left(  -\beta\Delta\phi_{i}+\Delta S_{i}^{\text{trc}}\right)
>1.\tag{16}%
\end{equation}

Without the steric term $\Delta S^{\text{trc}}$, this inequality is the
well-known Scharfetter-Gummel stability condition in semiconductor device
simulations \cite{SG69,S88}%
\begin{equation}
-\Delta\phi_{i}=-(\phi_{i+1}-\phi_{i})\leq\frac{2}{\beta}=\frac{2k_{B}T}%
{q}\text{ (for\ }\beta>0\text{)}\tag{17}%
\end{equation}
required to ensure that the FD equation (12) does not produce unphysical
approximations. Note that $q=2e$ for Ca$^{2+}$ yields an upper bound
$k_{B}T/e$ in (17), which is a half of that for Na$^{+}$ and means that if the
potential difference $-\Delta\phi_{i}$ between two adjacent points is greater
than $k_{B}T/e\approx25.7$ mV at room temperature, the resulting approximation
of the Ca$^{2+}$ flux $\mathbf{J}_{\text{Ca}^{2+}}$ in (9) may be completely
unphysical, although the same discretization of the Na$^{+}$ flux
$\mathbf{J}_{\text{Na}^{+}}$ may still be feasible. In other words, the FD
formula (12) is more unstable for Ca$^{2+}$ than for Na$^{+}$. In fact, if the
SG condition is violated, Newton's iteration for solving the coupled PNP
system of nonlinear algebraic equations is generally divergent. Of course, we
could reduce the mesh size $\Delta x$ so that the difference $-\Delta\phi_{i}$
is small enough to satisfy (17) at all grid points $i$. This would however
incur larger algebraic systems (and thus larger conditioning numbers of the
system) for which the computational cost would be more expensive. Even using
adaptive meshes that efficiently resolve internal or boundary layer regions
where $-\Delta\phi_{i}$ varies sharply, the primitive approximation (12) would
still diverge or show extremely slow convergence if the layer thickness is
very small \cite{CL03,CL05,CL08}.

The convergence and stability issues are further complicated by the steric
potential $S^{\text{trc}}$ in ion channel simulations if it is added to the FD
flux equation (12) as given in (13). From (16), we obtain a new SG condition
for ion channels%
\begin{equation}
-\beta\Delta\phi_{i}+\Delta S_{i}^{\text{trc}}\leq2\tag{18}%
\end{equation}
that will be a focal point in our numerical investigations in Section 5.

\textbf{\underline{Stabilization\textbf{.}}} To stabilize (12), we extend the
classical Scharfetter-Gummel approximation \cite{SG69} of the flux $J(x)$ to
include the steric potential such that
\begin{equation}
J_{i+\frac{1}{2}}=-\frac{D}{\Delta x}\left[  B(-t_{i})C_{i+1}-B(t_{i}%
)C_{i}\right] \tag{19}%
\end{equation}
where $t_{i}=\beta\Delta\phi_{i}-\Delta S_{i}^{\text{trc}}$ and $B(t)=\frac
{t}{e^{t}-1}$ is the Bernoulli function \cite{S88}. Eq. (19) is an exponential
fitting scheme for the concentration function $C(x)$ between the mesh points
$x_{i}$ and $x_{i+1}$ and is derived from the assumption that the flux $J$,
the local electric field $\frac{-d\phi}{dx}$, and the local steric field
$\frac{dS^{\text{trc}}}{dx}$ are all constant in this subinterval, i.e.,
\begin{equation}
\frac{J}{D}=\frac{-dC(x)}{dx}-kC(x)\text{, for all }x\in(x_{i}\text{, }%
x_{i+1})\text{,}\tag{20}%
\end{equation}
where $k=\beta\frac{d\phi}{dx}-\frac{dS^{\text{trc}}}{dx}$. Solving this
ordinary differential equation (ODE) with a boundary condition $C_{i}$ or
$C_{i+1}$ yields the well-known Goldman-Hodgkin-Katz flux equation in ion
channels \cite{H01}, which is exactly the same as that in (19).

The generalized Scharfetter-Gummel method for (11) is thus%
\begin{align}
\frac{dJ(x_{i})}{dx}  &  \approx\frac{J_{i+\frac{1}{2}}-J_{i-\frac{1}{2}}%
}{\Delta x}=\frac{a_{i-1}C_{i-1}+a_{i}C_{i}+a_{i+1}C_{i+1}}{\Delta x^{2}%
}=0\tag{21}\\
J_{i-\frac{1}{2}}  &  =\frac{-D}{\Delta x}\left[  B(-t_{i-1})C_{i}%
-B(t_{i-1})C_{i-1}\right] \nonumber\\
J_{i+\frac{1}{2}}  &  =\frac{-D}{\Delta x}\left[  B(-t_{i})C_{i+1}%
-B(t_{i})C_{i}\right] \nonumber\\
t_{i}  &  =\beta\Delta\phi_{i}-\Delta S_{i}^{\text{trc}}\text{, }B(t)=\frac
{t}{e^{t}-1}\nonumber\\
a_{i-1}  &  =-B(t_{i-1})\text{, }a_{i}=B(-t_{i-1})+B(t_{i})\text{, }%
a_{i+1}=-B(-t_{i})\text{. }\nonumber
\end{align}
The SG method is \textit{optimal} in the sense that it integrates the ODE (20)
\textit{exactly} at \textit{every} grid point with a suitable boundary
condition \cite{MR83}. Therefore, the SG method can resolve sharp layers very
accurately \cite{MR83} and hence needs few grid points to obtain tolerable
approximations when compared with the primitive FD method. Moreover, the exact
solution of (20) for the concentration function $C(x)$ yields an exact flux
$J(x)$. Consequently, the SG method is current-preserving, which is
particularly important in nonequilibrium systems, where the current is
possibly the most relevant physical property of interest \cite{BM89}. It is
difficult to overstate the importance of the current preserving feature and it
must be emphasized for workers coming from fluid mechanics that preserving
current has a significance quite beyond the preserving of flux in uncharged
systems. The electric field is so strong that the tiniest error in preserving
current, i.e., the tiniest deviation from Maxwell's equations, produces huge
effects. The third paragraph of Feynman's lectures on electrodynamics makes
this point unforgettable \cite{FL63}. Thus, the consequences of a seemingly
small error in preserving the flow of charge are dramatically larger than the
consequences of the same error in preserving the flux of mass.\bigskip

\begin{center}
\textbf{IV. SMIB-SG AND MULTISCALE METHODS\bigskip}
\end{center}

To test the PNPF theory and verify the numerical methods developed in this
paper, we consider the GA channel with a real protein structure and a
simplified molecular model of L-type calcium channels. The main difference
between these two channels is that the GA channel has a more rigid and less
negatively charged pore with about 2 \AA \ in radius whereas the Ca$^{2+}$
channel has a flexible and higher negatively charged binding site with radius
varying from 1 \AA \ to 2.5 \AA . The GA channel is also much longer (22 \AA ,
see below) than the selectivity filter of the L type calcium channel (10
\AA \ \cite{BV09}). Consequently, the GA channel is only cation selective
whereas the Ca$^{2+}$ channel is exquisitely Ca$^{2+}$ selective. The steric
potential is a key component of PNPF to properly describe this important
difference in selectivity along with the size effect of water (1.4 \AA \ in
radius). We use two different treatments of water that yield two different
steric potentials and size effects. \bigskip

\begin{center}
\textbf{A. The SMIB-SG Method for Gramicidin A Channel\bigskip}
\end{center}

Fig. 1(a) is a top view of the GA channel downloaded from the Protein Data
Bank \cite{B02}. A 2D cross section of the 3D simulation domain of the channel
embedded in a membrane is sketched in Fig. 1(b), where the biomolecular domain
$\Omega_{m}$ is composed of the channel protein and the membrane and the
solvent domain $\Omega_{s}$ consists of extracellular (upper), channel pore
(central), and intracellular (lower) regions. Particle species are indexed by
1, 2, and 3, for K$^{+}$, Cl$^{-}$, and H$_{2}$O with radii $a_{1}%
=a_{\text{K}^{+}}$, $a_{2}=a_{\text{Cl}^{-}}$, and $a_{3}=a_{\text{H}%
_{2}\text{O}}$ given in Table I.

The SMIB method is an advanced method to treat singularities of protein
charges and molecular surfaces \cite{ZC11,GY07,L13}. In SMIB, the electric
potential generated by the protein charges ($q_{j}\delta(\mathbf{r}%
-\mathbf{r}_{j})$ in (8)) is modeled as a sum of an analytical Green function
$\phi^{\ast}$ in infinite space and the Laplace potential $\phi^{0}$ in the
biomolecular domain $\Omega_{m}$ with boundary values of $\phi^{\ast}$ on
$\partial\Omega_{m}$. The combined potential then defines an electric field
$-\nabla(\phi^{\ast}+\phi^{0})$ that acts on ions and water in the solvent
domain $\Omega_{s}$ from the molecular surface $\partial\Omega_{m} $. The
total potential $\phi$ of all charged objects (ions, atomic charges, and
polarized water) is then calculated by solving Eqs. (6) and (7) with the SMIB
method for Eq. (7) across the interface $\partial\Omega_{m}$ of dielectric
solvent $\Omega_{s}$ and molecular $\Omega_{m}$ domains.%
\begin{figure}[ptb]%
\centering
\includegraphics[
height=2.578in,
width=4.5593in
]%
{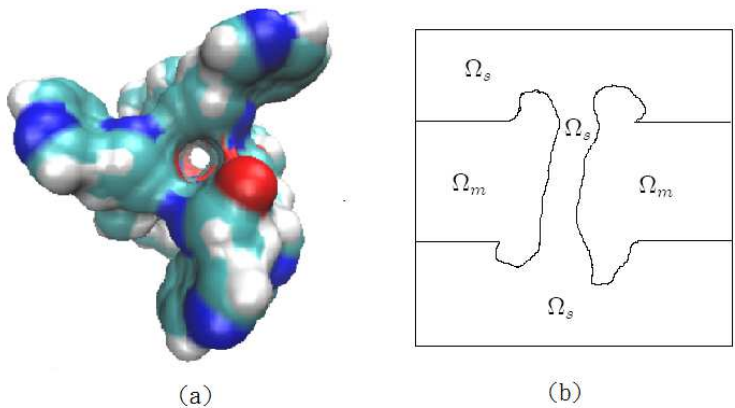}%
\caption{(Color online.) (a) Top view of the gramicidin A channel. (b) A cross
section of 3D simulation domain for the GA channel. The channel is placed in a
cubic box with the length of each side being 40 \AA \ and the thickness of the
membrane being 24 \AA .}%
\end{figure}

\begin{center}
$%
\begin{tabular}
[c]{c|c|c|c}%
\multicolumn{4}{c}{TABLE I. Notations and Physical Constants}\\\hline
Symbol & Meaning & \ Value & \ Unit\\\hline
\multicolumn{1}{l|}{$k_{B}$} & \multicolumn{1}{|l|}{Boltzmann constant} &
\multicolumn{1}{|l|}{$1.38\times10^{-23}$} & \multicolumn{1}{|l}{J/K}\\
\multicolumn{1}{l|}{$T$} & \multicolumn{1}{|l|}{temperature} &
\multicolumn{1}{|l|}{$298.15$} & \multicolumn{1}{|l}{K}\\
\multicolumn{1}{l|}{$e$} & \multicolumn{1}{|l|}{proton charge} &
\multicolumn{1}{|l|}{$1.602\times10^{-19}$} & \multicolumn{1}{|l}{C}\\
\multicolumn{1}{l|}{$\epsilon_{0}$} & \multicolumn{1}{|l|}{permittivity of
vacuum} & \multicolumn{1}{|l|}{$8.85\times10^{-14}$} &
\multicolumn{1}{|l}{F/cm}\\
\multicolumn{1}{l|}{$\epsilon_{\text{w}}$} & \multicolumn{1}{|l|}{water
dielectric constant} & \multicolumn{1}{|l|}{80 or $78.5$} &
\multicolumn{1}{|l}{}\\
\multicolumn{1}{l|}{$\epsilon_{\text{m}}$} & \multicolumn{1}{|l|}{protein
dielectric constant} & \multicolumn{1}{|l|}{2} & \multicolumn{1}{|l}{}\\
$\widehat{\epsilon}=\epsilon_{s}(1-l_{c}^{2}\nabla^{2})$ &
\multicolumn{1}{|l|}{dielectric operator, $\epsilon_{s}=\epsilon_{\text{w}%
}\epsilon_{0}$} & \multicolumn{1}{|l|}{in Eq. (6)} & \multicolumn{1}{|l}{F/cm}%
\\
\multicolumn{1}{l|}{$\widetilde{\epsilon}(\mathbf{r})\approx\widehat{\epsilon
}$} & \multicolumn{1}{|l|}{dielectric function} & \multicolumn{1}{|l|}{in Eqs.
(6), (31)} & \multicolumn{1}{|l}{}\\
\multicolumn{1}{l|}{$a_{\text{Na}^{+}}$, $a_{\text{Ca}^{2+}}$} &
\multicolumn{1}{|l|}{particle radii} & \multicolumn{1}{|l|}{$0.95$, $0.99$} &
\multicolumn{1}{|l}{\AA }\\
\multicolumn{1}{l|}{$a_{\text{K}^{+}}$, $a_{\text{Cl}^{-}}$, $a_{\text{H}%
_{2}\text{O}}$} & \multicolumn{1}{|l|}{particle radii} &
\multicolumn{1}{|l|}{1.33, $1.81$, $1.4$} & \multicolumn{1}{|l}{\AA }\\
\multicolumn{1}{l|}{$l_{c}$} & \multicolumn{1}{|l|}{correlation length} &
\multicolumn{1}{|l|}{1.2$a_{\text{K}^{+}}$ or 2$a_{\text{Ca}^{2+}}$} &
\multicolumn{1}{|l}{\AA }\\
\multicolumn{1}{l|}{$D_{\text{K}^{+}}^{\text{B}}$} &
\multicolumn{1}{|l|}{K$^{+}$ diffusion coefficient} &
\multicolumn{1}{|l|}{$1.96\times10^{-5}$} & \multicolumn{1}{|l}{cm$^{2}/$s}\\
\multicolumn{1}{l|}{$D_{\text{Na}^{+}}^{\text{B}}$} &
\multicolumn{1}{|l|}{Na$^{+}$ diffusion coefficient} &
\multicolumn{1}{|l|}{$1.334\times10^{-5}$} & \multicolumn{1}{|l}{cm$^{2}/$s}\\
\multicolumn{1}{l|}{$D_{\text{Ca}^{2+}}^{\text{B}}$} &
\multicolumn{1}{|l|}{$\text{Ca}^{2+}$ diffusion coefficient} &
\multicolumn{1}{|l|}{$0.792\times10^{-5}$} & \multicolumn{1}{|l}{cm$^{2}/$s}\\
\multicolumn{1}{l|}{$D_{\text{Cl}^{-}}^{\text{B}}$} &
\multicolumn{1}{|l|}{$\text{Cl}^{-}$ diffusion coefficient} &
\multicolumn{1}{|l|}{$2.032\times10^{-5}$} & \multicolumn{1}{|l}{cm$^{2}/$s}\\
\multicolumn{1}{l|}{$D_{\text{H}_{2}\text{O}}^{\text{B}}$} &
\multicolumn{1}{|l|}{$\text{H}_{2}\text{O}$ diffusion coefficient} &
\multicolumn{1}{|l|}{$2.3\times10^{-5}$} & \multicolumn{1}{|l}{cm$^{2}/$s}\\
\multicolumn{1}{l|}{$V_{\text{i,o}}$} & \multicolumn{1}{|l|}{inside (outside)
voltage} & \multicolumn{1}{|l|}{} & \multicolumn{1}{|l}{V}\\\hline
\end{tabular}
\bigskip$
\end{center}

The molecular surface $\partial\Omega_{m}$ as depicted in Fig. 1(a) is
generated by rolling a probe ball (water molecule) with radius 1.4 \AA \ over
a total of 554 spherical atoms in the GA protein \cite{SR73}. In SMIB, the
molecular surface is not fixed and is adaptively determined by the grid size
so that the interface point is always in the middle of neighboring grid
points. The resulting surface is thus free of geometric singularities. We
refer to \cite{L13} for more details about the SMIB and surface generation methods.

The NP equation (9) is then solved by the SG method for each particle species
$i$ once $\phi$ is known. An iterative process of solving PF1 (6), PF2 (7),
and NP equations is repeated again until convergent approximations of
$\phi(\mathbf{r})$ and $C_{i}(\mathbf{r})$ are found at all grid points. As
noted above, convergence of this kind of iterative process is in general not
guaranteed and must be checked at all grid points. We propose the following
nonlinear iteration algorithm for the PNPF system (6), (7), and (9) using SMIB
and SG methods:

\textit{Nonlinear Iteration Algorithm 1:}

\begin{enumerate}
\item[1.] Solve the Laplace equation $-\nabla^{2}\phi^{0}(\mathbf{r})=0$ in
$\Omega_{m}$ for the potential $\phi^{0}(\mathbf{r})$ once for all with
$\phi^{0}(\mathbf{r})=\phi^{\ast}(\mathbf{r})=\sum_{j=1}^{Q}q_{j}%
/(4\pi\epsilon_{m}\epsilon_{0}\left\vert \mathbf{r-r}_{j}\right\vert )$ on
$\partial\Omega_{m}$.

\item[2.] Solve the Poisson equation $-\nabla\cdot\left(  \epsilon\nabla
\phi(\mathbf{r})\right)  =0$ in $\Omega$ at equilibrium for the initial
potential $\phi^{\text{Old}}(\mathbf{r})$ with $\phi^{\text{Old}}=0$ on
$\partial\Omega$ and the jump condition $\left[  \epsilon\nabla\phi
^{\text{Old}}\cdot\mathbf{n}\right]  =-\epsilon_{m}\epsilon_{0}\nabla
(\phi^{\ast}+\phi^{0})\cdot\mathbf{n}$ on $\partial\Omega_{m}$, where $\left[
u\right]  $ denotes the jump function across $\partial\Omega_{m}$ \cite{L13}.

\item[3.] Solve the PF1 $\epsilon_{s}\left(  l_{c}^{2}\nabla^{2}-1\right)
\Psi(\mathbf{r})=\sum_{i=1}^{K}q_{i}C_{i}^{\text{Old}}(\mathbf{r})$ in
$\Omega_{s}$ for $\Psi^{\text{New}}(\mathbf{r})$ with $\nabla\Psi^{\text{New}%
}\cdot\mathbf{n}=0$ on $\partial\Omega_{m}$, $\Psi^{\text{New}}=0$ on
$\partial\Omega$, and the Fermi distribution $C_{i}^{\text{Old}}%
(\mathbf{r})=C_{i}^{\text{B}}\exp\left(  -\beta_{i}\phi^{\text{Old}%
}(\mathbf{r})+S^{\text{trc}}(\mathbf{r})\right)  $,\ $S^{\text{trc}%
}(\mathbf{r})=\ln\frac{\Gamma^{\text{Old}}(\mathbf{r)}}{\Gamma^{\text{B}}}$,
$\Gamma(\mathbf{r)}=1-\sum_{j=1}^{K+1}v_{j}C_{j}^{\text{Old}}(\mathbf{r})$.

\item[4.] Solve the linearized PF2 $-\nabla\cdot\left(  \epsilon\nabla
\phi(\mathbf{r})\right)  +\rho^{\prime}(\phi^{\text{Old}})\phi(\mathbf{r}%
)=-\epsilon\Psi^{\text{New}}+\rho^{\prime}(\phi^{\text{Old}})\phi^{\text{Old}%
}$ at equilibrium for the next potential $\phi^{\text{New}}(\mathbf{r})$ with
the same jump and boundary conditions in Step 2. Here $\rho^{\prime}(\phi)$
denotes the derivative of the charge density functional $\rho(\phi)=\sum
_{i=1}^{K}q_{i}C_{i}^{\text{B}}\exp\left(  -\beta_{i}\phi+S^{\text{trc}%
}\right)  $ in $\Omega_{s}$ with respect to $\phi$.

\item[5.] Assign $\phi^{\text{Old}}=\omega_{\text{PF}}\phi^{\text{Old}%
}+(1-\omega_{\text{PF}})\phi^{\text{New}}$ with a suitable relaxation
parameter $\omega_{\text{PF}}$ and go to Step 3 if the error $\left\Vert
\phi^{\text{New}}-\phi^{\text{Old}}\right\Vert _{\infty}$ in the infinity norm
is larger than a preset tolerance, else go to Step 6.

\item[6.] Solve the steady state NP equation $-\nabla\cdot\mathbf{J}%
_{i}(\mathbf{r})=0$ in $\Omega_{s}$ at nonequilibrium for $C_{i}^{\text{New}%
}(\mathbf{r})$ and all $i=1,\cdots,K+1$ with $\mathbf{J}_{i}(\mathbf{r}%
)=-D_{i}\left[  \nabla C_{i}(\mathbf{r})+\beta_{i}C_{i}(\mathbf{r})\right.  $
$\left.  \nabla\phi^{\text{Old}}(\mathbf{r})-C_{i}(\mathbf{r})\nabla
S^{\text{trc}}(\mathbf{r})\right]  $,\ $S^{\text{trc}}(\mathbf{r})=\ln
\frac{\Gamma^{\text{Old}}(\mathbf{r)}}{\Gamma^{\text{B}}}$, $C_{i}%
^{\text{New}}(\mathbf{r})=0$ on $\partial\Omega$, and $\mathbf{J}%
_{i}(\mathbf{r})\cdot\mathbf{n}=0$ on $\partial\Omega_{m}$.

\item[7.] Solve the PF1 for $\Psi^{\text{New}}$ as in Step 3 with
$C_{i}^{\text{New}}$ in place of $C_{i}^{\text{Old}}$.

\item[8.] Solve the PF2 $-\nabla\cdot\left(  \epsilon\nabla\phi(\mathbf{r}%
)\right)  =-\epsilon\Psi^{\text{New}}$ (without linearization) at
nonequilibrium for $\phi^{\text{New}}$.

\item[9.] Assign $\phi^{\text{Old}}=\omega_{\text{PNPF}}\phi^{\text{Old}%
}+(1-\omega_{\text{PNPF}})\phi^{\text{New}}$ with a suitable relaxation
parameter $\omega_{\text{PNPF}}$ and go to Step 6 if $\left\Vert
\phi^{\text{New}}-\phi^{\text{Old}}\right\Vert _{\infty}$ is larger than a
preset error tolerance, else stop.
\end{enumerate}

This is an SOR-like iteration algorithm modified from that in \cite{ZC11}. The
modifications include the additional solution processes at equilibrium in
Steps 2, 3, and 4, the extra PF1 in Steps 3 and 7, Newton's linearization for
PF2 in Step 4, two relaxation parameters $\omega_{\text{PF}}$ and
$\omega_{\text{PNPF}}$ in Step 5 and 9 with $0<\omega_{\text{PF}}$,
$\omega_{\text{PNPF}}$ $<1$ (under relaxation), and the extra water NP
equation $-\nabla\cdot\mathbf{J}_{K+1}(\mathbf{r})=0$ in Step 6. The stability
and convergence rate are controlled by these two parameters. If the parameter
is close to zero, we will have more stable iteration but slower convergence.
The correlation length $l_{c}$ and the Fermi distribution (or the steric
potential $S^{\text{trc}}$) in Step 3 signify the difference between the
classical PNP and advanced PNPF models. The stability and convergence are
further complicated by these two physical properties for which a continuation
method may be needed by introducing two stepping parameters $\lambda_{c}$ and
$\lambda_{S}$ such that $\lambda_{c}l_{c}$ and $\lambda_{S}S^{\text{trc}}$ are
gradually increased from $\lambda_{c}=\lambda_{S}=0 $ to $\lambda_{c}%
=\lambda_{S}=1$ \cite{L13}.

The water NP equation is not considered in previous PNP models and plays an
essential role not only for numerical stability but also for physical validity
because the void volume fraction $\Gamma(\mathbf{r)}=1-\sum_{j=1}^{K+1}%
v_{j}C_{j}(\mathbf{r})$ in Step 3 at any location $\mathbf{r}$ in the solvent
domain $\Omega_{s}$ needs to be carefully checked in each iteration. Numerical
errors in approximating the concentration functions $C_{j}(\mathbf{r})$ for
any particle species $j=1,\cdots,K+1$ could easily lead to an unphysical void
fraction $\Gamma(\mathbf{r)}<0$ at some $\mathbf{r}$. Water molecules
automatically adjust themselves in the PNPF model and move together with all
ions in the system as the above iteration process converges to a stable and
correct state, although the net water flow through the channel may be zero.
Moreover, the water NP equation also dynamically determines the variable
permittivity $\widetilde{\epsilon}(\mathbf{r})\epsilon_{0}=\epsilon
_{s}/(1+\eta(\mathbf{r})/\rho(\mathbf{r}))$ from the bath to the pore and thus
automatically adjusts dielectric forces on ions along the channel pathway.
These dielectric forces can have a decisive effect on biologically important
conductance \cite{NH03} and on selectivity. For example, Na$^{+}$ vs. K$^{+}$
selectivity in Na$^{+}$ channels is only found when the dielectric function is
handled in more detail \cite{BB02,BN07}.

However, this SMIB-SG method and previous PNP methods suffer from a major
difficulty in ion channel simulations. Those methods have difficulty in
dealing with the essential property of selectivity, which of course is
different in different types of channels with different structures. The L-type
calcium channel selects Ca$^{2+}$ over Na$^{+}$ of similar size and a
potassium channel selects K$^{+}$ over Na$^{+}$ of the same charge. The
following method is proposed to overcome this difficulty.\bigskip

\begin{center}
\textbf{B. A Multiscale Method for Calcium Channel\bigskip}
\end{center}

Calcium channels have not yet been crystallized and so we use the
Lipkind-Fozzard molecular model \cite{LF01} of L-type calcium channels in
which the EEEE locus (four glutamate side chains modeled by 8 O$^{1/2-}$ ions)
forms a high-affinity Ca$^{2+}$ binding site that is essential to Ca$^{2+}$
selectivity, blockage, and permeation. Fig. 2(a) illustrates the binding site
and the EEEE locus, where 3 Ca$^{2+}$ are shown in violet, 8 O$^{1/2-}$ in
red, 2 H$_{2}$O in white and red. Fig. 2(b) is a cross section of a simplified
3D channel geometry for the present work, where the central circle denotes the
binding site, the other four circles denote the side view of 8 O$^{1/2-}$
ions, $\Omega_{s}$ is the solvent domain consisting of two baths and the
channel pore including the binding domain $\overline{\Omega}_{\text{Bind}}$,
$\Omega_{m}$ is the biomolecular domain with the boundary $\partial\Omega_{m}%
$, and $\partial\Omega$ is the outside and inside bath boundary. Fig. 3 is a
sketch of the binding site and O$^{1/2-}$ ions, where $d_{O}^{Ca}$ is the
distance between the center of a binding Ca$^{2+}$ ion and the center $c_{j}$
of any O$^{1/2-}$, and $A$ is any point on the surface of the site. In our
model, the 8 O$^{1/2-}$ ions are not contained in the solvent domain
$\Omega_{s}$. Particle species are indexed by 1, 2, 3, and 4 for Na$^{+}$,
Ca$^{2+}$, Cl$^{-}$, and H$_{2}$O, respectively.%
\begin{figure}[ptb]%
\centering
\includegraphics[
height=2.5763in,
width=5.5988in
]%
{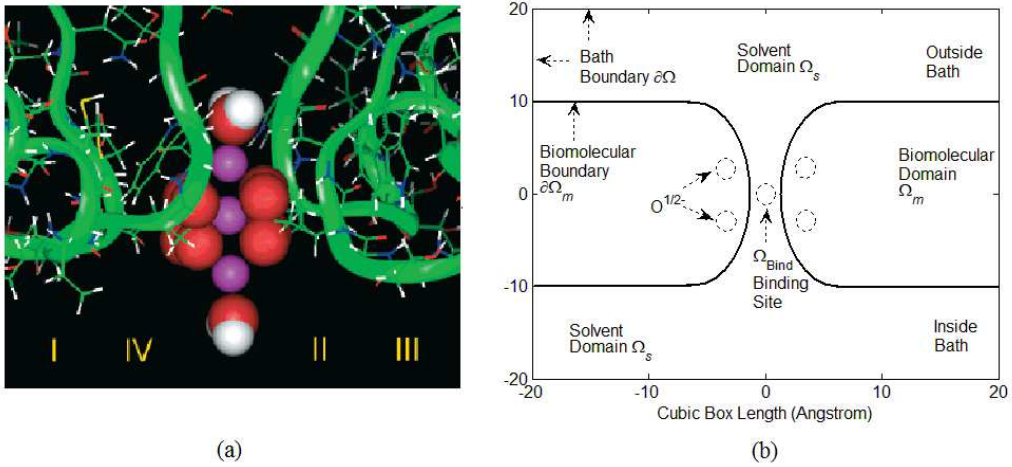}%
\caption{(Color online.) (a) The Lipkind-Fozzard pore model, where 3 Ca$^{2+}$
are shown in violet, 8 O$^{1/2-}$ in red, 2 H$_{2}$O in white and red.
Reprinted with permission from (G. M. Lipkind and H. A. Fozzard, Biochem.
\textbf{40}, 6786 (2001)). Copyright (2001) American Chemical Society. (b) A
simplified Ca channel geometry in a cubic box with baths, pore, and binding
site. The solvent region $\Omega_{s}$ consists of two baths and the channel
pore. The binding site $\overline{\Omega}_{\text{Bind}}$ is contained in
$\Omega_{s}$ but the O$^{1/2-}$ ions are not in $\Omega_{s}$. The outside and
inside bath boundary is denoted by $\partial\Omega$.}%
\end{figure}
\begin{figure}[ptb]%
\centering
\includegraphics[
height=2.4344in,
width=4.088in
]%
{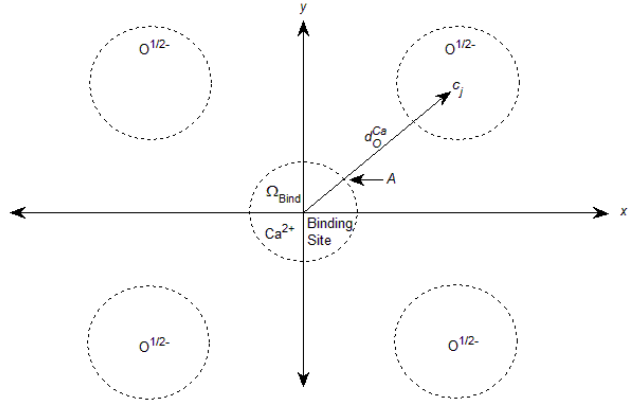}%
\caption{The binding distance between the center of the binding Ca$^{2+}$ ion
and the center $c_{j}$ of the $j^{\text{th}}$ O$^{1/2-}$ ion is denoted by
$d_{O}^{Ca}$ for $j=1,\cdots,8$. $A$ is any point on the surface of the
binding ion.}%
\end{figure}

In \cite{LE14}, we proposed an algebraic model for calculating the electrical
potential $\overline{\phi}_{b}$ and the steric potential $\overline{S}%
_{b}^{\text{trc}}$ in $\overline{\Omega}_{\text{Bind}}$ by using Coulomb's law
with the atomic structure of binding ion and atoms in a channel protein as
shown in Fig. 3, without solving the Poisson-Fermi equation (8) in
$\overline{\Omega}_{\text{Bind}}$. The volume of $\overline{\Omega
}_{\text{Bind}}$ is an unknown variable $v_{b}$ that changes with different
charges in the binding site. The algebraic model \cite{LE14} defined in
$\overline{\Omega}_{\text{Bind}}$ consists of the following equations
\begin{equation}
\left\{
\begin{array}
[c]{l}%
O_{1}^{b}=v_{b}C_{1}^{\text{B}}\exp(-\beta_{1}\overline{\phi}_{b}+\overline
{S}_{b}^{\text{trc}})\\
O_{2}^{b}=v_{b}C_{2}^{\text{B}}\exp(-\beta_{2}\overline{\phi}_{b}+\overline
{S}_{b}^{\text{trc}})\\
O_{4}^{b}=v_{b}C_{4}^{\text{B}}\exp(\overline{S}_{b}^{\text{trc}}),
\end{array}
\right. \tag{22}%
\end{equation}%
\begin{equation}
\overline{S}_{b}^{\text{trc}}=\ln\frac{v_{b}-v_{1}O_{1}^{b}-v_{2}O_{2}%
^{b}-v_{4}O_{4}^{b}}{v_{b}\Gamma^{\text{B}}},\tag{23}%
\end{equation}%
\begin{equation}
\frac{e}{4\pi\epsilon_{0}}\left(  \sum_{j=1}^{8}\frac{z_{\text{O}^{1/2-}}%
}{|c_{j}-A|}+\frac{O_{1}^{b}z_{\text{Na}^{+}}}{a_{\text{Na}^{+}}}+\frac
{O_{2}^{b}z_{\text{Ca}^{2+}}}{a_{\text{Ca}^{2+}}}\right)  =\overline{\phi}%
_{b},\tag{24}%
\end{equation}
where $O_{1}^{b}$, $O_{2}^{b}$, and $O_{4}^{b}$ denote the occupancy numbers
of Na$^{+}$, Ca$^{2+}$, and H$_{2}$O in $v_{b}$, respectively, $\overline
{\phi}_{b}$ and $\overline{S}_{b}^{\text{trc}}$ are average electrical and
steric potentials, and $|c_{j}-A|$ is the distance between $A$ and $c_{j}$ in
Fig. 3.

In this mean field, we allow $O_{1}^{b}$ and $O_{2}^{b}$ (and hence the total
charge $O_{1}^{b}ez_{\text{Na}^{+}}+O_{2}^{b}ez_{\text{Ca}^{2+}}$) to vary
continuously subject to the condition on their sum $O_{1}^{b}+O_{2}^{b}=1$ in
the binding volume $v_{b}$. Eqs. (22) and (23) uniquely determine the four
unknowns $v_{b}$, $O_{4}^{b}$, $\overline{\phi}_{b}$ and $\overline{S}%
_{b}^{\text{trc}}$ with $O_{1}^{b}$ and $O_{2}^{b}$ being given. Eq. (24)
uniquely determines the locations ($c_{j}$) of 8 O$^{1/2-}$ ions (and thus the
binding distance $d_{O}^{Ca}$ or $d_{O}^{Na}$ in Fig. 3) once $\overline{\phi
}_{b}$ is obtained. Note that the binding distance $d_{O}^{O_{1}^{b}%
Na+O_{2}^{b}Ca}$ (or $c_{j}$) changes continuously with varying $O_{1}^{b}$
and $O_{2}^{b}$ but $\overline{\phi}_{b}$ remains fixed, where the binding ion
$O_{1}^{b}Na+O_{2}^{b}Ca$ is a linear combination of Na$^{+}$ and Ca$^{2+}$.
Therefore, O$^{1/2-}$ ions are movable --- the protein is flexible in our
model --- as their locations $c_{j}$ changes with varying $O_{1}^{b}$ and
$O_{2}^{b}$.

For the half-blockage experimental condition \cite{AM84}%
\begin{equation}
\underset{\text{{\normalsize Experimental Data}}}{\underbrace{C_{\text{Na}%
^{+}}^{\text{B}}=C_{\text{1}}^{\text{B}}=32\text{ mM, }C_{\text{Ca}^{2+}%
}^{\text{B}}=C_{\text{2}}^{\text{B}}=0.9\text{ }\mu\text{M,}}}\tag{25}%
\end{equation}
we follow convention and assume relative occupancies of a filled channel,
$O_{1}^{b}=0.5$ and $O_{2}^{b}=0.5$, and thereby obtain $\overline{\phi}%
_{b}=-10.48$ $k_{B}T/e$, $\overline{S}_{b}^{\text{trc}}=-1.83$, and
$v_{b}=4.56$ \AA $^{3}$ \cite{LE14}. The binding experiments \cite{AM84} used
a fixed $C_{\text{Na}^{+}}^{\text{B}}=C_{\text{1}}^{\text{B}}=32$ mM and
various Ca$^{2+}$ bath concentrations $C_{\text{Ca}^{2+}}^{\text{B}%
}=C_{\text{2}}^{\text{B}}$ that imply different $O_{1}^{b}$ and $O_{2}^{b}$ of
Na$^{+}$ and Ca$^{2+}$ occupying the binding site. The occupancy numbers
$O_{1}^{b}$ and $O_{2}^{b}$ are determined by
\begin{equation}
\frac{O_{1}^{b}}{O_{2}^{b}}=\frac{1-O_{2}^{b}}{O_{2}^{b}}=\exp(-(\beta
_{1}-\beta_{2})\overline{\phi}_{b})\frac{C_{1}^{\text{B}}}{C_{2}^{\text{B}}%
},\tag{26}%
\end{equation}
where $\overline{\phi}_{b}$ was just obtained from the case of equal
occupancy. The occupancy ratio in (26) thus deviates from unity as
$C_{2}^{\text{B}}$ is varied along the horizontal axis of the binding curve
from its midpoint value $C_{2}^{\text{B}}=0.9$ $\mu$M as shown in Fig. 5 in
\cite{LE14}.

For nonequilibrium cases, the binding steric potential $\overline{S}%
_{b}^{\text{trc}}$ is assigned its equilibrium value in subsequent PNPF
calculations, i.e., the void fraction $\Gamma(\mathbf{r})$ in $\overline
{\Omega}_{\text{Bind}}$ is assumed to remain unchanged from equilibrium to
nonequilibrium. The electrical potential $\overline{\phi}_{b}$ will be
modified by the membrane potential $V_{\text{i}}-V_{\text{o}}$ \cite{KC99} and
then used as a Dirichlet type condition for the potential function
$\phi(\mathbf{r})$ in $\overline{\Omega}_{\text{Bind}}$. For this multiscale
method, the boundary conditions for the PF (8) and NP (9) equations are
\begin{equation}
\left\{
\begin{array}
[c]{l}%
\text{ }\phi(\mathbf{r})=\widetilde{\phi}_{b}(\mathbf{r})\text{ in }%
\overline{\Omega}_{\text{Bind}}\text{, }\phi(\mathbf{r})=V_{\text{o,i}}\text{
on }\partial\Omega\text{,}\\
C_{i}(\mathbf{r})=C_{i}^{\text{B}}\text{ on }\partial\Omega\text{,
}i=1,2,3,4,\\
\mathbf{J}_{i}(\mathbf{r})\cdot\mathbf{n}=0\text{ on }\partial\Omega
_{m}\text{.}%
\end{array}
\right. \tag{27}%
\end{equation}
Note that the electrostatic potential $\phi(\mathbf{r})$ is prescribed as a
Dirichlet function $\widetilde{\phi}_{b}(\mathbf{r})$ whose spatial average in
$\overline{\Omega}_{\text{Bind}}$ is the constant $\overline{\phi}_{b}$.
However, the binding domain $\overline{\Omega}_{\text{Bind}}$ is treated as an
interior domain instead of boundary domain for the NP equation (9).

If a condition on the boundary is used to solve the Poisson (or PF) equation
as in Algorithm 1, the resulting steric potential $\overline{S}_{b}%
^{\text{trc}}$ (as an output of $\phi(\mathbf{r})$ by (5)) may be incorrect in
$\overline{\Omega}_{\text{Bind}}$ because the atomic equations (23) and (24)
are not used. We do not have any differential equation for the steric function
$S^{\text{trc}}(\mathbf{r})$ for which appropriate boundary conditions near
$\overline{\Omega}_{\text{Bind}}$ can be imposed if a conventional method is
used. The methods proposed in this paper are still coarse approximations to
ion transport as the PNPF theory is in its early development. Nevertheless,
the theory provides many atomic properties such as (23) and (24) that have
been shown to be important for studying the binding mechanism in
Ca$_{\text{V}}$ channels \cite{LE14} and are also important for the transport
mechanism as shown in the next section. Incorporating atomic properties into
continuum models is a step forward to improve and refine the continuum theory.
We refer to \cite{LE14b,LE14} for more details of the algebraic model and its
extension to PNPF.

We summarize the PNPF solution process using the multiscale method as follows.

\textit{Nonlinear Iteration Algorithm 2:}

\begin{enumerate}
\item[1.] Solve (23) and (24) for $\overline{\phi}_{b}$ and $\overline{S}%
_{b}^{\text{trc}}$ in the binding site $\overline{\Omega}_{\text{Bind}}$ with
the experimental data (25).

\item[2.] Choose any linear interpolation $\overline{\phi}^{\text{Old}}$ (an
initial guess potential profile) that links the binding potential $\phi_{A}$
to the zero potential at each bath boundary for the potential function
$\phi(\mathbf{r})$.

\item[3.] Solve the PB equation $-\nabla\cdot\left(  \epsilon\nabla
\phi(\mathbf{r})\right)  =\rho(\overline{\phi}^{\text{Old}})=\sum_{i=1}%
^{K}q_{i}\overline{C}_{i}^{\text{Old}}$ at equilibrium for $\phi^{\text{Old}}$
with the Boltzmann distribution $\overline{C}_{i}^{\text{Old}}=C_{i}%
^{\text{B}}\exp\left(  -\beta_{i}\overline{\phi}^{\text{Old}}\right)  $.
Compute the initial concentrations $C_{i}^{\text{Old}}=C_{i}^{\text{B}}%
\exp\left(  -\beta_{i}\phi^{\text{Old}}\right)  $.

\item[4.] Solve the PF1 $\epsilon_{s}\left(  l_{c}^{2}\nabla^{2}-1\right)
\Psi(\mathbf{r})=\sum_{i=1}^{K}q_{i}C_{i}^{\text{Old}}(\mathbf{r})$ in
$\Omega_{s}$ for $\Psi^{\text{New}}(\mathbf{r})$ with the same conditions as
in Algorithm 1.

\item[5.] Solve the linearized PF2 $-\nabla\cdot\left(  \epsilon\nabla
\phi(\mathbf{r})\right)  +\rho^{\prime}(\phi^{\text{Old}})\phi(\mathbf{r}%
)=-\epsilon\Psi^{\text{New}}+\rho^{\prime}(\phi^{\text{Old}})\phi^{\text{Old}%
}$ in $\Omega_{s}$ at nonequilibrium for $\phi^{\text{New}}$ with the
conditions in (27).

\item[6.] Solve the NP equation $-\nabla\cdot\mathbf{J}_{i}(\mathbf{r})=0$ in
$\Omega_{s}$ at nonequilibrium for $C_{i}^{\text{New}}(\mathbf{r})$ and all
$i=1,\cdots,K+1$ with the same conditions as in Algorithm 1.

\item[7.] Go to Step 4 if $\left\Vert \phi^{\text{New}}-\phi^{\text{Old}%
}\right\Vert _{\infty}$ or $\left\Vert C_{i}^{\text{New}}-C_{i}^{\text{Old}%
}\right\Vert _{\infty}$ is larger than a preset error tolerance, else stop.
\end{enumerate}

We do not need to solve the Poisson equation in the biomolecular domain
$\Omega_{m}$ that contains the singular charges of 8 O$^{1/2-}$, since the
effect of these charges on potentials has been included in the integral
constraint we apply to the binding potential $\overline{\phi}_{b}$ in (24).
Consequently, we do not have to deal with the delta function in (8) and the
potential jump conditions on $\partial\Omega_{m}$ as used in Algorithm 1. The
absence of jump conditions makes the approximation of PF1 and PF2 more
accurate since numerical methods for handling the jump conditions across
molecular surfaces with singular cusps are subtle, complex, and thus prone to
error \cite{GY07,L13}. Moreover, the SOR-like scheme is not needed for this
iteration. Application of the multiscale method to the NCX structure
\cite{LL12} is briefly discussed in \cite{LE14}. It will be interesting to
apply the method to the celebrated KcsA potassium channel \cite{DM98} and to
recent structures of TRPV1 \cite{LC13} and Ca$_{\text{V}}$Ab channels
\cite{TC14}.

\begin{center}
\textbf{V. NUMERICAL RESULTS}
\end{center}

The main purpose of this work is to present numerical methods that are
suitable for continuum simulations of ion transport in different types of
biological ion channels with particular interests in treating the excluded
volume effect of all particles and the dynamical effect of water molecules.
Numerical methods are validated for accuracy with exact solutions of the PNP
model for the GA channel. Numerical results of the PNPF model for both GA and
calcium channels are all verified with experimental data.

\begin{center}
\textbf{A. Gramicidin A Channel}
\end{center}

The Scharfetter-Gummel method (21) is first validated with the following exact
solutions for the PNP model \cite{ZC11}
\begin{equation}
\phi(\mathbf{r})=\left\{
\begin{array}
[c]{l}%
\cos x\cos y\cos z\text{,\ \ }\mathbf{r}=(x,y,z)\in\Omega_{m}\text{,}\\
\cos x\cos y\cos z\text{,\ \ }\mathbf{r}\in\Omega_{s}\text{,}%
\end{array}
\right. \tag{28}%
\end{equation}%
\begin{equation}
C_{1}(\mathbf{r})=\left\{
\begin{array}
[c]{l}%
0\text{ in }\Omega_{m}\text{,}\\
0.2\cos x\cos y\cos z+0.3\text{ in }\Omega_{s}\text{,}%
\end{array}
\right. \tag{29}%
\end{equation}%
\begin{equation}
C_{2}(\mathbf{r})=\left\{
\begin{array}
[c]{l}%
0\text{\ in }\Omega_{m}\text{,}\\
0.1\cos x\cos y\cos z+0.3\text{ in }\Omega_{s}.
\end{array}
\right. \tag{30}%
\end{equation}
Note that the right hand side of the Poisson equation in Algorithm 1 is not
zero as the exact solution (28) has been imposed and the Green function
$\phi^{\ast}(\mathbf{r})=\sum_{j=1}^{Q}q_{j}/\left(  4\pi\epsilon_{m}\right.
$ $\left.  \left\vert \mathbf{r-r}_{j}\right\vert \right)  $ is only used in
the jump condition on the molecular surface $\partial\Omega_{m}$, where the
coordinates $\mathbf{r}_{j}$ of the atoms in the GA channel protein are
provided in the Protein Data Bank \cite{B02}, the protein charge $q_{j}$ and
the radius of each atom $j$ are obtained by the PDB2PQR software \cite{DC07},
and the total number of atoms is $Q=554$. The optimal convergence (second)
order, i.e. $O(h^{2})$, of the SMIB method for the nonlinear Poisson-Boltzmann
equation has been confirmed in \cite{L13}. The need for such validation has
been pointed out before \cite{HC01}. It is easy to mistake convergence for
accuracy in systems of PNP like equations \cite{J95}.

For a full nonlinear PNP system (without steric, correlation, and water NP
effects) using Algorithm 1, Table II shows that the optimal convergence order
has been achieved for all PNP equations as well by using the SMIB method for
the Poisson equation and the primitive FD method (12) for the NP equations in
the nonlinear iteration process. In the table, errors are measured in the
$L_{\infty}$ norm. For example, $0.0927=\max_{ijk}\left\vert \phi(x_{i}%
,y_{j},z_{k})-\phi_{ijk}\right\vert $, where $\phi_{ijk}$ is the FD
approximation of the Poisson equation and $\phi(x_{i},y_{j},z_{k}) $ is the
exact value evaluated by (28) at the grid point $(x_{i},y_{j},z_{k}) $ with
the mesh size $h=1$ \AA . The error tolerance for both linear solver and
nonlinear iteration was set to 10$^{-6}$. All errors and orders (Ord) of
convergence in Table II are similar to those in \cite{ZC11}, showing that the
SMIB method in \cite{L13} is comparable to the original MIB \cite{ZC11}.

When the primitive FD method is replaced by the SG method (21) for NP
equations, it is surprising that the preset error tolerance 10$^{-6}$ was
satisfied by all SG approximations $\phi_{ijk}$, $C_{1}^{ijk}$, and
$C_{2}^{ijk}$ at all grid points for all mesh sizes as shown in Table III.
Errors in Table III are much more smaller than those in Table II. This
demonstrates that the SG is an optimal (exponential fitting) method to
discretize the NP equation as implied by the exact analysis of the ODE (20),
since all solution functions in (28)-(30) are very smooth so that the
assumptions made in (20) are valid. It only took 2 nonlinear iterations and
about 1 hour and 8 minutes on a laptop computer with 2.6 GHz Intel CPU to
reach the error tolerance for the case of $h=0.25$ \AA . The corresponding
matrix size is about 4.2 millions. The maximum potential difference
$\Delta\phi_{i}$ between any two adjacent grid points for the most coarse case
($h=1$ \AA ) is -1.045 (not shown), which satisfies the SG condition (18).
This illustrates why the convergence has been achieved by the primitive FD
without SG as shown in Table II.\bigskip

\begin{center}%
\begin{tabular}
[c]{c|cccccc|ccc}%
\multicolumn{7}{c|}{TABLE II. Errors in $L_{\infty}$ norm by FD} &
\multicolumn{3}{c}{TABLE III. Errors by SG}\\\hline
$h$ ($\text{\AA )}$ & P & Ord & NP1 & Ord & NP2 & Ord & \ P \  & \ NP1 \  &
NP2\\\hline
1.00 & 0.0927 &  & 0.0505 &  & 0.0211 &  & \ 10$^{-6}$ \  & 10$^{-6}$ &
10$^{-6}$\\
0.50 & 0.0245 & 1.91 & 0.0076 & 2.73 & 0.0042 & 2.33 & 10$^{-6}$ & 10$^{-6}$ &
10$^{-6}$\\
0.25 & 0.0060 & 2.03 & 0.0019 & 2.00 & 0.0010 & 2.07 & 10$^{-6}$ & 10$^{-6}$ &
10$^{-6}$\\\hline
\end{tabular}
\bigskip
\end{center}

We now study full PNPF (with steric, correlation, and water NP effects)
simulations of the GA channel using the SMIB and SG methods. Fig. 4 is a
comparison of the I-V curves obtained by PNPF (lines) and the experimental
data (symbols) from Cole et al. \cite{CF02} with bath K$^{+}$ and Cl$^{-}$
concentrations $C^{\text{B}}=0.1$, 0.2, 0.5, 1, and 2 M and membrane
potentials $\Delta V=V_{\text{i}}-V_{\text{o}}=0$, 50, 100, 150, and 200 mV.
The PNPF currents in pico ampere (pA) were obtained with only one adjustable
parameter, namely, the reduction parameter $\theta$ in the pore diffusion
coefficients $\theta D_{i}^{\text{B}}$ for all particle species, while all
physical parameters in Table I were kept fixed throughout simulations. This
kind of reduction parameter has been used in all previous PNP papers and is
necessary in continuum simulations when compared with MD, BD, or experimental
data because there is abundant qualitative evidence that the diffusion
coefficient in channels is much smaller than in bulk, but quantitative
estimates are not available, as well described by Gillespie in \cite{G08}
including Appendix and supporting material. In principle, all experimental
data can be fitted by adjusting this parameter. For the PNPF currents at all
$C^{\text{B}}$ and $\Delta V$ in Fig. 4, we chose $\theta=1/4.7$ which agrees
with the range 1/3 to 1/10 obtained by many MD simulations of various channel
models \cite{AK00,MC03,SS98}, indicating that the steric, correlation, and
water NP properties have made PNPF simulations more closer (realistic) to MD
simulations than previous PNP simulations for which the parameter $\theta$
differs from MD values by an order to several orders of magnitude \cite{AK00}.%
\begin{figure}[ptb]%
\centering
\includegraphics[
height=2.5771in,
width=4.4114in
]%
{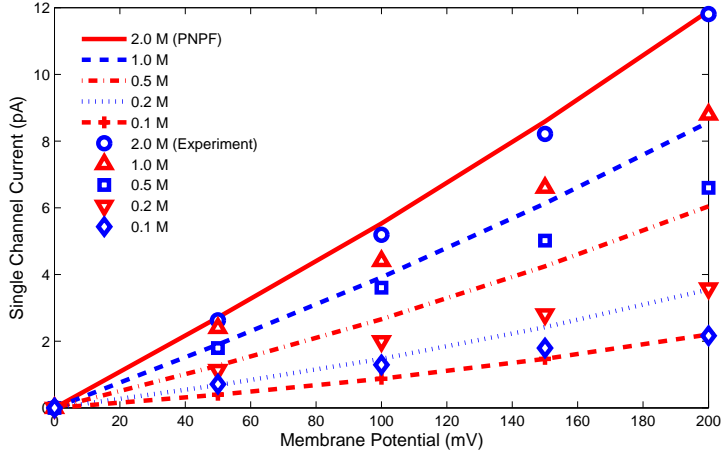}%
\caption{A comparison of PNPF (lines) and experimental \cite{CF02} (symbols)
I-V results with bath K$^{+}$ and Cl$^{-}$ concentrations $C^{\text{B}}=0.1$,
0.2, 0.5, 1, 2 M and membrane potentials $\Delta V=0$, 50, 100, 150, 200 mV.}%
\end{figure}

Furthermore, PNPF can also provide more physical properties that have not been
observed by previous PNP models such as the variation of electric permittivity
(dielectric function $\widetilde{\epsilon}(\mathbf{r})$ in Fig. 5) and water
density ($C_{\text{H}_{2}\text{O}}(\mathbf{r})$ in Fig. 6) from bath to
channel pore. Together with the electric ($\phi(\mathbf{r})$ in Fig. 7) and
steric ($S^{\text{trc}}(\mathbf{r})$ in Fig. 8) potentials, K$^{+}$ ions (in
Fig. 9) are subject not only to the electric field $-\nabla\phi(\mathbf{r})$
but also to the steric (entropic) field $\nabla S^{\text{trc}}(\mathbf{r})$ as
described in Eq. (10). These fields change with the variations of water
density, other ion concentrations, voids $\Gamma(\mathbf{r})$, and dielectric
function $\widetilde{\epsilon}(\mathbf{r})$ at any location $\mathbf{r}$ in
the solvent domain $\Omega_{s}$. For example, the magnitude of electric fields
modified by the dielectric function $\widetilde{\epsilon}(\mathbf{r})$ can be
as large as $(80-50)/80=37.5\%$ of that by the constant permittivity
80$\epsilon_{0}$ for the bath condition $C^{\text{B}}=2$ M with the membrane
potential 200 mV as shown in Fig. 5. The dielectric function in
$\widetilde{\epsilon}(\mathbf{r})\epsilon_{0}$ was calculated by
\begin{equation}
\widetilde{\epsilon}(\mathbf{r})=\epsilon_{m}+C_{\text{H}_{2}\text{O}%
}(\mathbf{r})(\epsilon_{\text{w}}-\epsilon_{m})/C_{\text{H}_{2}\text{O}%
}^{\text{B}}\tag{31}%
\end{equation}
using the water density function $C_{\text{H}_{2}\text{O}}(\mathbf{r})$ as
proposed in \cite{LM08}. The protein is most negatively charged around $z=13$
\AA , where the pore is very narrow (about 1.6 \AA \ in radius) so that it is
most crowded (most negative $S^{\text{trc}}(\mathbf{r})=\ln\frac
{\Gamma(\mathbf{r})}{\Gamma^{B}}$ in Fig. 8) there. The size effect of all
particle species is clearly manifested by the steric function $S^{\text{trc}%
}(\mathbf{r})$ in PNPF. These results provide one of the most comprehensive
simulations on ion transport in real proteins using continuum models that we
know of.%
\begin{figure}[ptb]%
\centering
\includegraphics[
height=2.5771in,
width=4.4114in
]%
{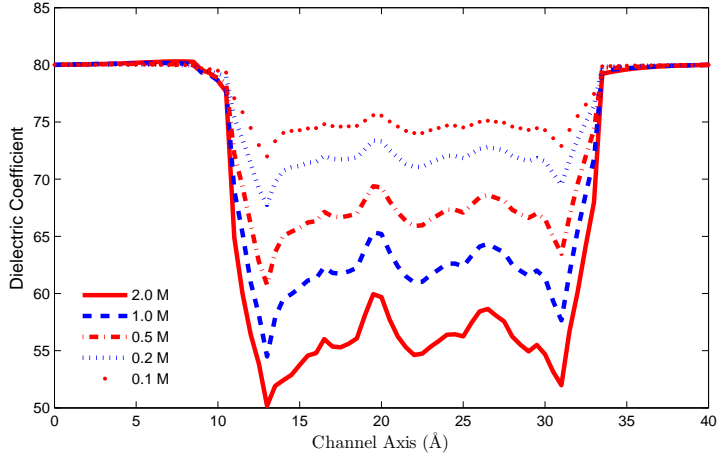}%
\caption{The averaged dielectric function $\protect\widetilde{\epsilon
}(\mathbf{r})$ profiles at each cross section along the pore axis with
$C^{\text{B}}=$ 0.1, 0.2, 0.5, 1, 2 M and $\Delta V=200$ mV. Figs. 5.3 -- 5.6
are obtained with the same averaging method, $C^{\text{B}}$, and $\Delta V$.}%
\end{figure}
\begin{figure}[ptb]%
\centering
\includegraphics[
height=2.5771in,
width=4.4114in
]%
{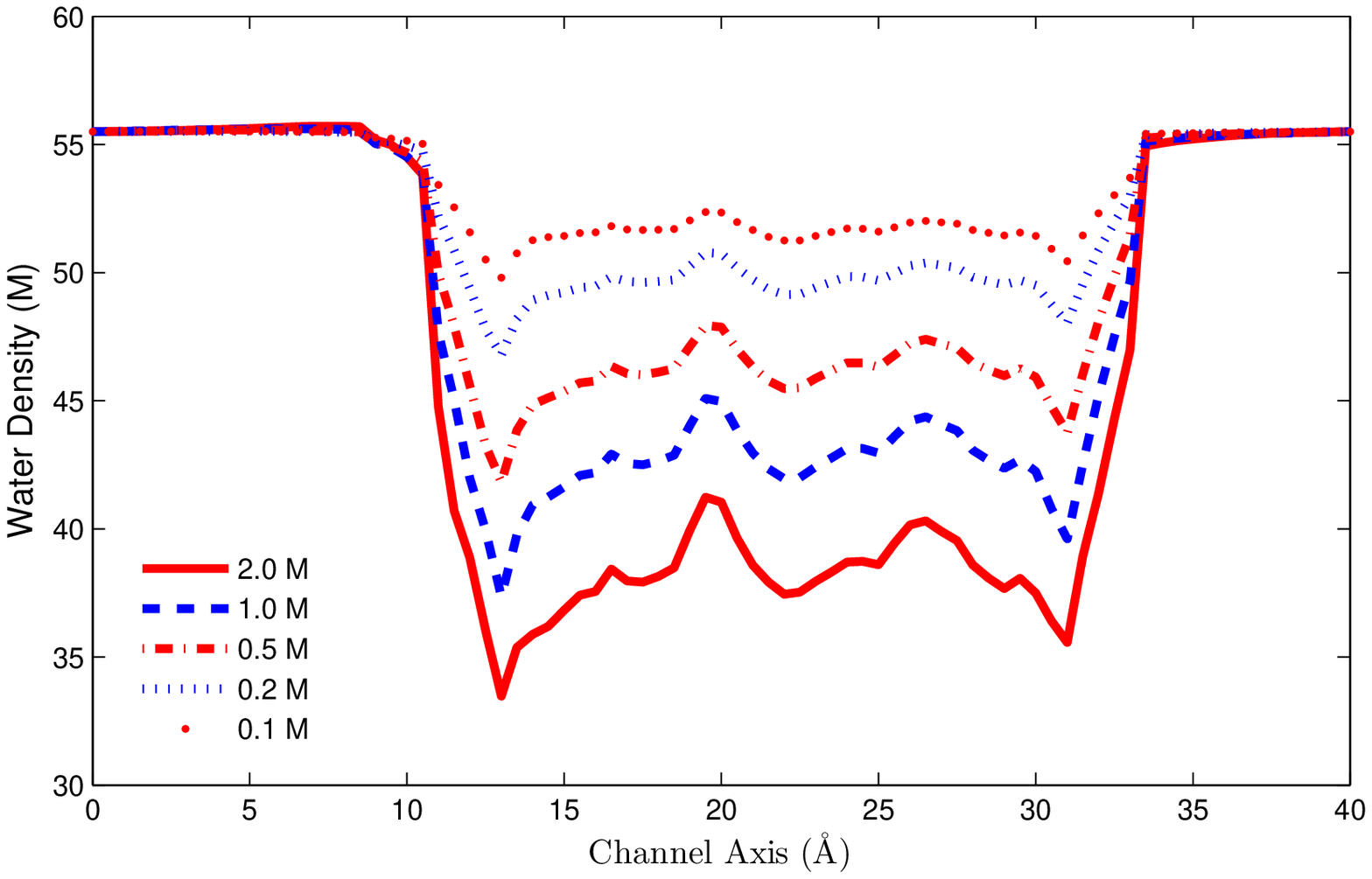}%
\caption{The averaged water density $C_{\text{H}_{2}\text{O}}(\mathbf{r})$
profiles.}%
\end{figure}
\begin{figure}[ptb]%
\centering
\includegraphics[
height=2.5771in,
width=4.4114in
]%
{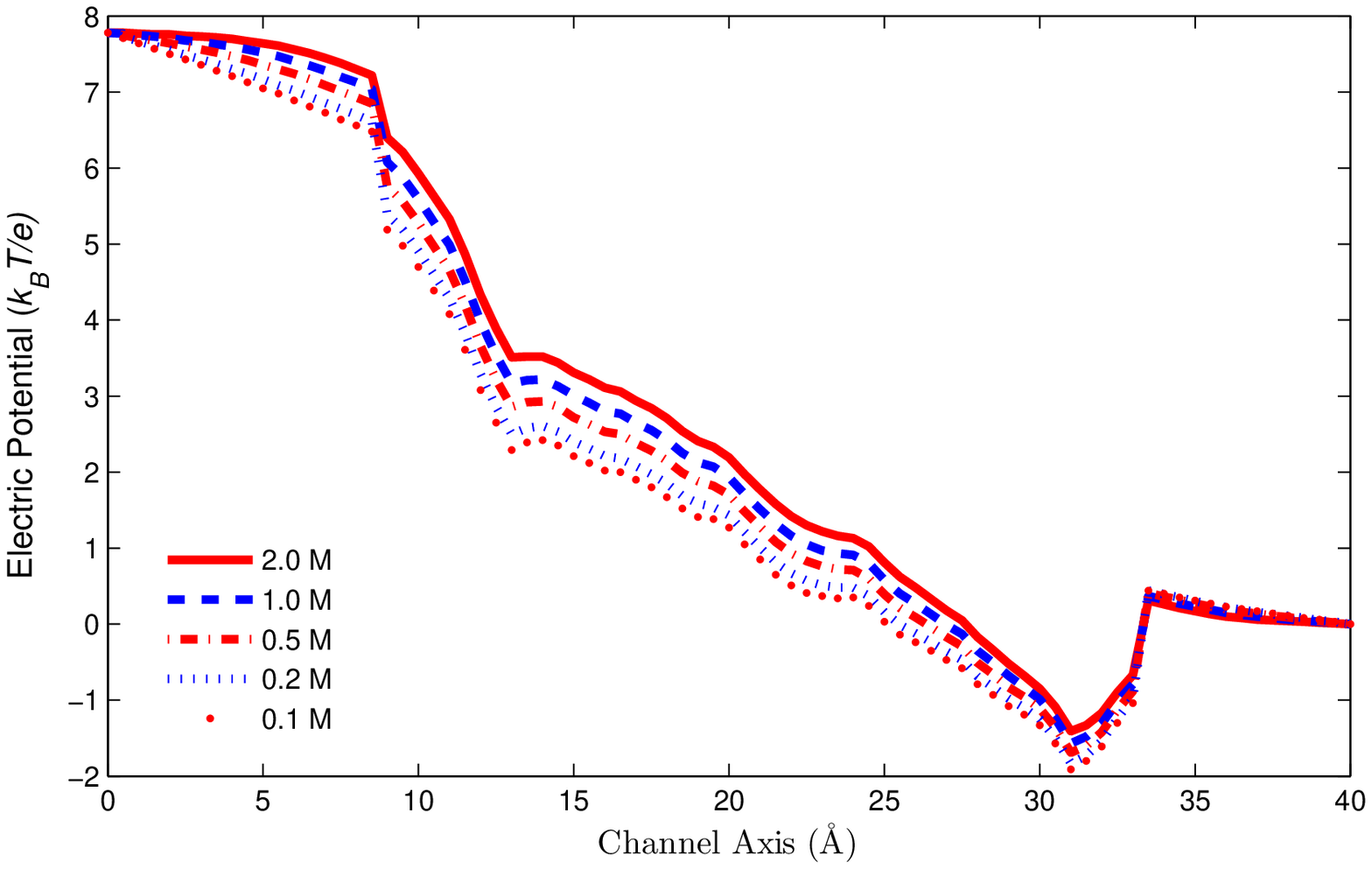}%
\caption{The averaged electric potential $\phi(\mathbf{r})$ profiles.}%
\end{figure}
\begin{figure}[ptb]%
\centering
\includegraphics[
height=2.5771in,
width=4.4114in
]%
{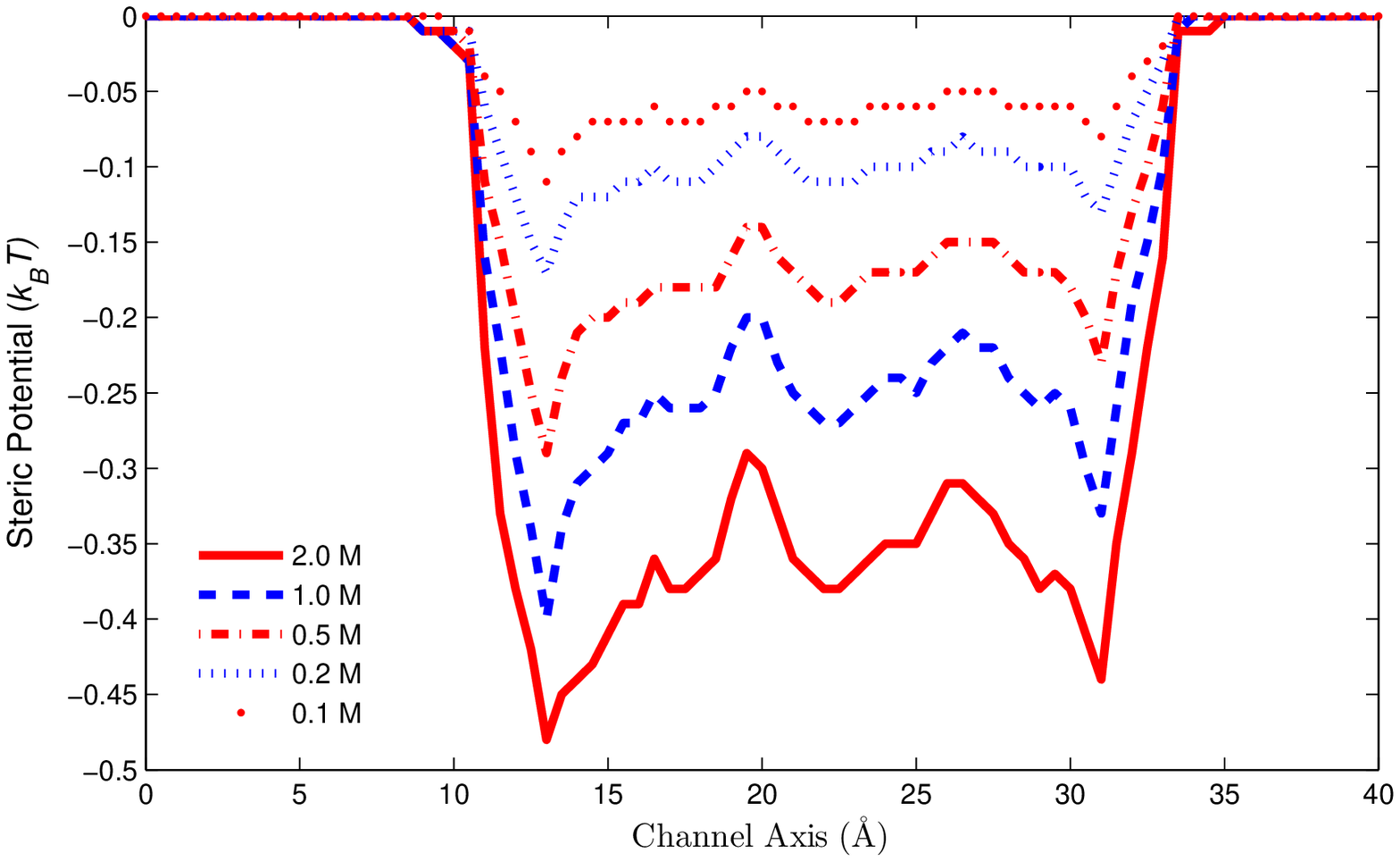}%
\caption{The averaged steric potential $S^{\text{trc}}(\mathbf{r})$ profiles.
}%
\end{figure}
\begin{figure}[ptb]%
\centering
\includegraphics[
height=2.5771in,
width=4.4114in
]%
{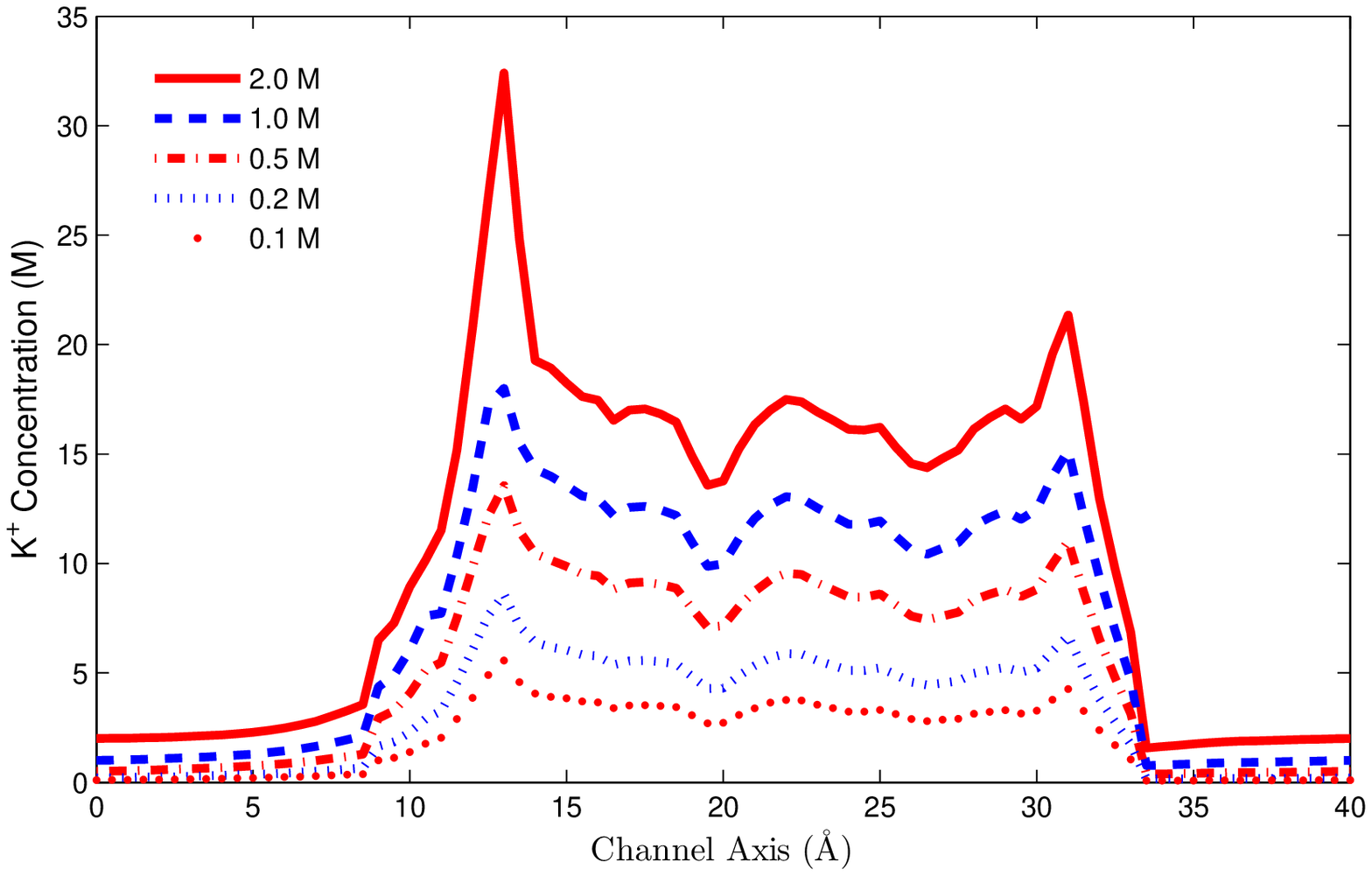}%
\caption{The averaged K$^{+}$ concentration $C_{\text{K}^{+}}(\mathbf{r})$
profiles.}%
\end{figure}

The incompressibility of water and the mass conservation are important
physical properties that can be used to further verify continuum results. MD
simulations have shown that the GA channel can be occupied by two $K^{+}$ ions
at moderately high concentration \cite{RP95,R02}. In Fig. 10(a), we observe
that a total of 8 particles (water molecules plus $K^{+}$ ions) in the channel
pore is conserved by PNPF but not by PNP as [KCl] increases from 0 to 2 M. The
pore volume is determined by a length of 22 \AA \ (from 11 to 33 \AA \ in the
channel axis in Fig. 9) and radii varying from 1.466 to 2.343 \AA \ along the
axis (not shown). The PNPF water density profiles in Fig. 6 show that water
molecules adjust self-consistently their configurations to accommodate $K^{+}$
ions (Fig. 9) in the two binding sites near the mouths of the channel as [KCl]
increases. The complementary profiles of water and $K^{+}$ in Figs. 6 and 9
illustrate a continuum picture of six water molecules separating two $K^{+}$
ions in single file \cite{R02}. Figs. 10(a) and 10(b) also illustrate the
saturation of ions and currents, respectively, as [KCl] increases. Note that
PNP yields less $K^{+}$ ions (and hence currents) since the constant
permittivity $\epsilon=$ $\epsilon_{\text{w}}\epsilon_{0}$ in Eq. (8) for PNP
(with $l_{c}=0$) is larger than the variable permittivity $\widetilde{\epsilon
}(\mathbf{r})\epsilon_{0}$ obtained by Eq. (31) for PNPF (with $l_{c}\neq0$)
as shown in Fig. 5, i.e., larger $\epsilon$ results in smaller charge density
$\rho$ (fewer ions) for the same $\phi$.

Therefore, the mass conservation and saturation results in Fig. 10 and the
PNPF current results in Fig. 4 with the MD compatible parameter $\theta=1/4.7$
appear to justify the approximation formula (31) for calculating the variable
$\widetilde{\epsilon}(\mathbf{r})$ that is an output for illustration. We
emphasize that in our treatment, unlike most treatments of channels,
dielectric and polarization effects are operators that are outputs of the
calculations. They are not assumed as constants. The polarization effects of
water are actually approximated by the dielectric operator $\widehat{\epsilon
}=\epsilon_{s}(1-l_{c}^{2}\nabla^{2})$ in Eq. (8) not by $\widetilde{\epsilon
}(\mathbf{r})$. Obviously, the polarization effects (or equivalently the
correlation effects of ions in PNPF) play a crucial role in very narrow
channels that are more challenging to describe by classical PNP models or even
by all-atom MD simulations as the current MD force fields do not include the
electronic polarization effects \cite{BK06,BK09}. Of course, the single-file
picture by PNPF is still far from that by MD \cite{R02} due to inevitable
averaging effects of numerous atoms in the system. Nevertheless, the
electrostatic, steric, and dielectric fields produced by PNPF may improve MD
as well as continuum simulations in future studies.%
\begin{figure}[ptb]%
\centering
\includegraphics[
height=2.5771in,
width=4.4114in
]%
{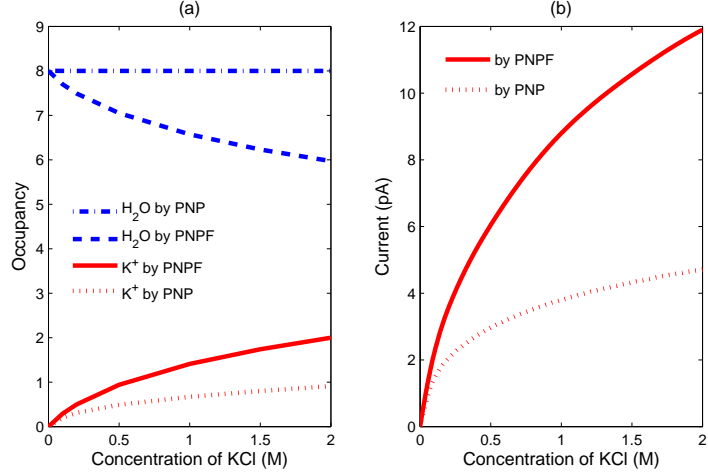}%
\caption{(a) Occupancy of H$_{2}$O and K$^{+}$ in the GA channel pore by PNPF
and PNP as [KCl] increases from 0 to 2 M. The total number of H$_{2}$O and
K$^{+}$ in the pore is 8 \cite{RP95}, which is conserved by PNPF but not by
PNP (without steric and correlation effects). (b) Currents by PNPF and PNP at
$\Delta V=200$ mV. PNP underestimates the currents.}%
\end{figure}

We make a final remark about the nonlinear iteration method for GA
simulations. The two relaxation parameters of the SOR-like scheme in Algorithm
1 were set to $\omega_{\text{PF}}=0.3$ and $\omega_{\text{PNPF}}=0.5$ for all
above results. The number of iterations for each PNPF I-V data point in Fig. 4
is given in Table IV. We do not need to solve NP equations when $\Delta V=0$.
Iterations for solving PF1 and PF2 in Steps 3 and 4 in Algorithm 1 increase
with increasing bath concentrations as shown in the table. Iterations for
solving K$^{+}$, Cl$^{-}$, and H$_{2}$O NP equations and then PF1 and PF2 in
Steps 6, 7, 8 are all about 22 for $C^{\text{B}}=0.1$ to 1 M when $\Delta
V\neq0$. These numbers are more steady and less than those in \cite{ZC11} (see
Table 7 in that paper). For $C^{\text{B}}=2$ M, iterations increase with
increasing $\Delta V$ as those in \cite{ZC11}. Note that the relaxation
parameter was set to different values for the first 3 steps in \cite{ZC11}
whereas $\omega_{\text{PF}}$ and $\omega_{\text{PNPF}}$ were fixed throughout here.

\begin{center}%
\begin{tabular}
[c]{c|ccccc}%
\multicolumn{6}{c}{TABLE IV. SOR Iterations}\\\hline
$C^{\text{B}}\backslash\Delta V$ & 0mV & 50 & 100 & 150 & 200\\\hline
\multicolumn{1}{l|}{\ 0.1M} & 22 & 22 & 22 & 22 & 22\\
\multicolumn{1}{l|}{\ 0.2} & 30 & 22 & 22 & 22 & 22\\
\multicolumn{1}{l|}{\ 0.5} & 45 & 21 & 22 & 22 & 22\\
\multicolumn{1}{l|}{\ 1.0} & 61 & 21 & 21 & 21 & 21\\
\multicolumn{1}{l|}{\ 2.0} & 78 & 34 & 38 & 43 & 49\\\hline
\end{tabular}
\bigskip

\textbf{B. Calcium Channel\bigskip}
\end{center}

The calcium channel operates very delicately in physiological and experimental
conditions as it shifts its exquisitely tuned conductance from Na$^{+}$-flow,
to Na$^{+}$-blockage, and to Ca$^{2+}$-flow when bath Ca$^{2+} $ varies from
trace to high concentrations. In \cite{AM84}, 19 extracellular solutions and 3
intracellular solutions were studied experimentally, in which the range of
[Ca$^{2+}$]$_{\text{o}}$ is 10$^{8}$-fold from $10^{-10.3} $ to $10^{-2}$ M.

PNPF results are in accord with the experimental data in \cite{AM84} as shown
in Fig. 11 under only the same salt conditions of NaCl and CaCl$_{2}$ in pure
water, without considering other bulk salts in experimental solutions. With
[$\text{Na}^{+}$]$_{\text{i}}=$ [$\text{Na}^{+}$]$_{\text{o}}=32$ mM and
[Ca$^{2+}$]$_{\text{i}}=0$, the membrane potential is fixed at $-20$ mV
($V_{\text{o}}=0$ and $V_{\text{i}}$ $=-20$ mV) throughout, as assumed in Fig.
11 of \cite{AM84} for all single-channel currents (in femto ampere fA)
recorded in the experiment. The small circles in Fig. 11 denote the
experimental currents from Fig. 11 of \cite{AM84} and the plus signs denote
the total currents calculated by PNPF, where the partial Ca$^{2+}$ and
Na$^{+}$ currents are denoted by the solid and dotted line, respectively.
These two ionic currents show the anomalous fraction effect of the channel at
nonequilibrium. The reduction parameter in $\theta D_{i}$ was set to
$\theta=0.1$ and all physical parameters in Table I were kept fixed
throughout.%
\begin{figure}[ptb]%
\centering
\includegraphics[
height=2.5771in,
width=4.4114in
]%
{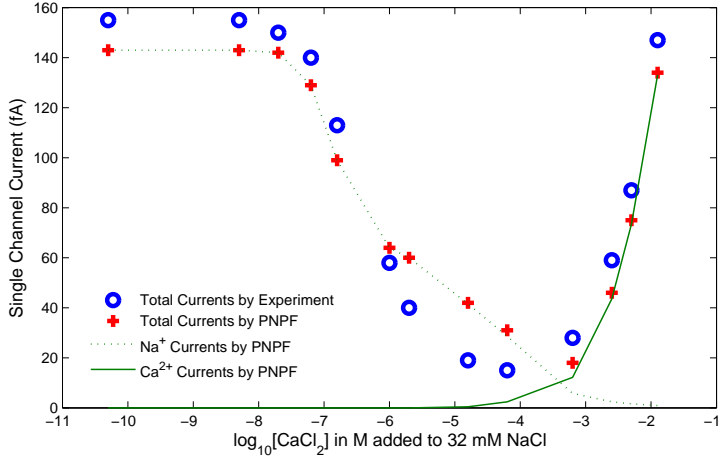}%
\caption{Single channel inward current in femto ampere (fA) plotted as a
function of $\log_{10}$[Ca$^{2+}$]$_{\text{o}}$. Experimental data of
\cite{AM84} are marked by small circles and PNPF data are denoted by the plus
sign and lines.}%
\end{figure}

Solution profiles of calcium channel are quite different from those of GA
channel as shown in Figs. 12 (dielectric function $\widetilde{\epsilon
}(\mathbf{r})$), 13 (water density $C_{\text{H}_{2}\text{O}}(\mathbf{r})$), 14
(electric potential $\phi(\mathbf{r})$), 15 (steric potential $S^{\text{trc}%
}(\mathbf{r})k_{B}T$), and 16 (scaled flux density$\left\vert \mathbf{J}%
_{\text{Ca}^{2+}}(\mathbf{r})\right\vert $). Fig. 12 displays the combined
effects of correlation, polarization, and screening in this highly
inhomogeneous electrolyte by means of the dielectric function
$\widetilde{\epsilon}(\mathbf{r})$ in which water (Fig. 13) plays a more
profound role than that of GA channel as the water density is dramatically
reduced from 55.5 M in the bath to almost 0 in the binding site when
[Ca$^{2+}$]$_{\text{o}}=10^{-2}$ M.

Water is not allowed to occupy the binding site because Ca$^{2+}$ occupies it
in this bath condition and the 8 O$^{1/2-}$ ions in the EEEE locus are
electrically attracted toward the binding Ca$^{2+}$ as illustrated by Fig.
2(a). The EEEE locus is very hydrophobic in this condition. Without using the
atomic properties of water and ions \textit{inside} the solvent domain
$\Omega_{s}$ like those in (22)-(24), continuum models are not likely to
produce results like Figs. 12 and 13. Mathematically, the Dirichlet condition
$\phi(\mathbf{r})=\widetilde{\phi}_{b}(\mathbf{r})$ in the interior of
$\Omega_{s}$, i.e. $\overline{\Omega}_{\text{Bind}}\subset$ $\Omega_{s}$ in
(27), is crucially important to connect the continuum Poisson-Fermi model (4)
in $\Omega_{s}\backslash\overline{\Omega}_{\text{Bind}}$ to the molecular
(Coulomb) model (22)-(24) in $\overline{\Omega}_{\text{Bind}}$. From the
binding formula (24), the pore radius is enlarged by the binding Na$^{+}$ when
[Ca$^{2+}$]$_{\text{o}}$ decreases from $10^{-2}$ to $10^{-7.2}$ M, i.e.,
Na$^{+}$ occupancy ($O_{1}$ in (26)) increases. The enlarged radius allows
more space for water molecules in the channel pore as shown in Fig. 13.%
\begin{figure}[ptb]%
\centering
\includegraphics[
height=2.5771in,
width=4.4114in
]%
{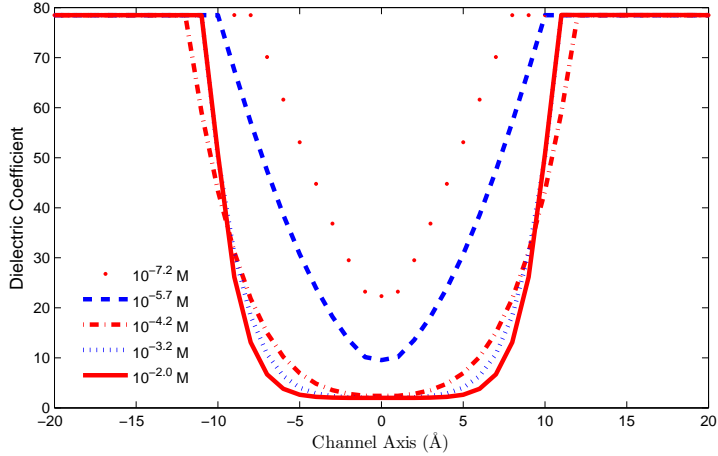}%
\caption{The averaged dielectric function $\protect\widetilde{\epsilon
}(\mathbf{r})$ profiles at each cross section along the pore axis for various
[Ca$^{2+}$]$_{\text{o}}$ ranging from 10$^{-7.2}$ M to 10$^{-2}$ M. All the
following figures are obtained with the same averaging method and the same
range of [Ca$^{2+}$]$_{\text{o}}$.}%
\end{figure}
\begin{figure}[ptb]%
\centering
\includegraphics[
height=2.5771in,
width=4.4114in
]%
{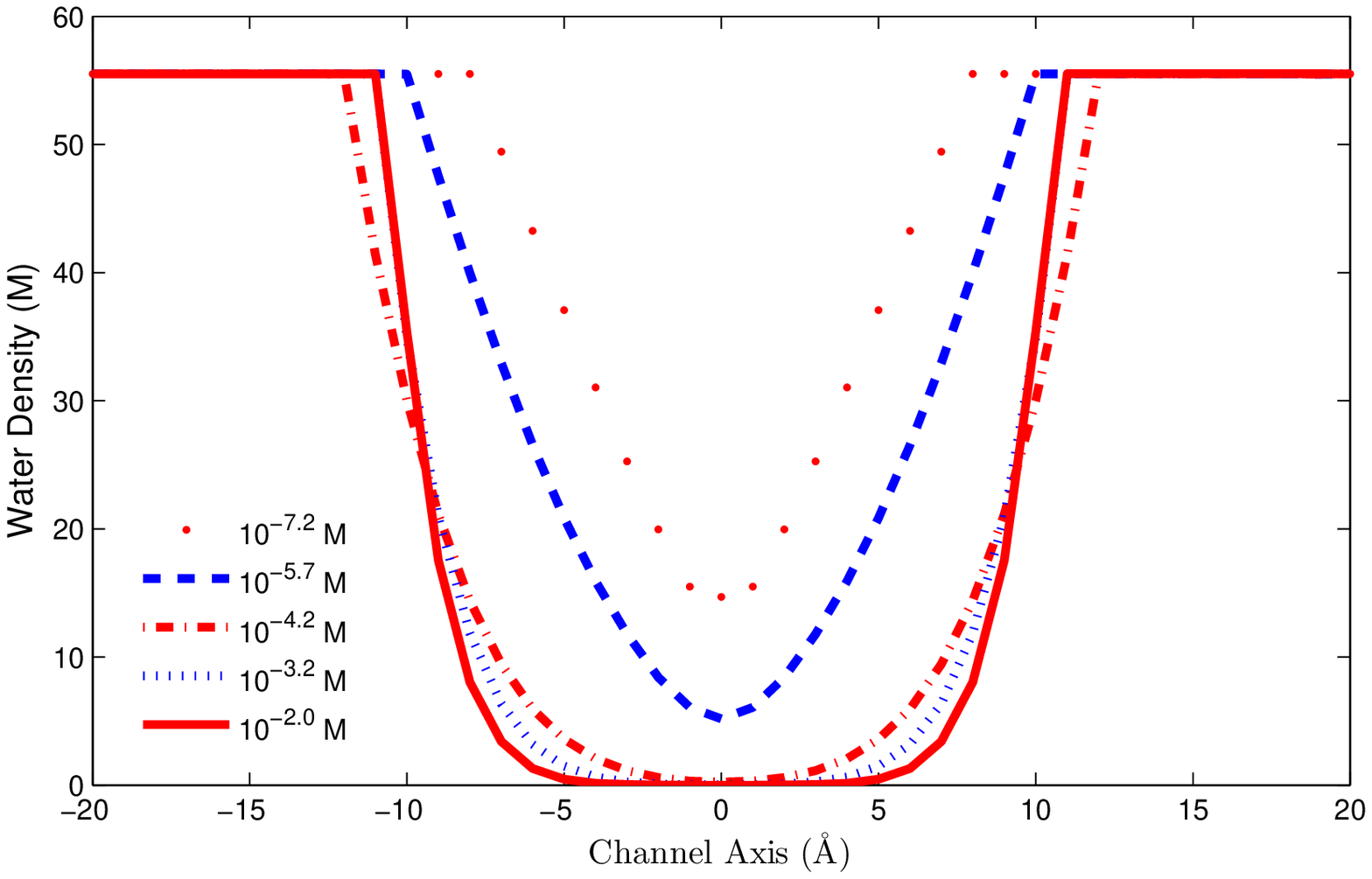}%
\caption{The averaged water density $C_{\text{H}_{2}\text{O}}(\mathbf{r})$
profiles.}%
\end{figure}
\begin{figure}[ptb]%
\centering
\includegraphics[
height=2.5771in,
width=4.4114in
]%
{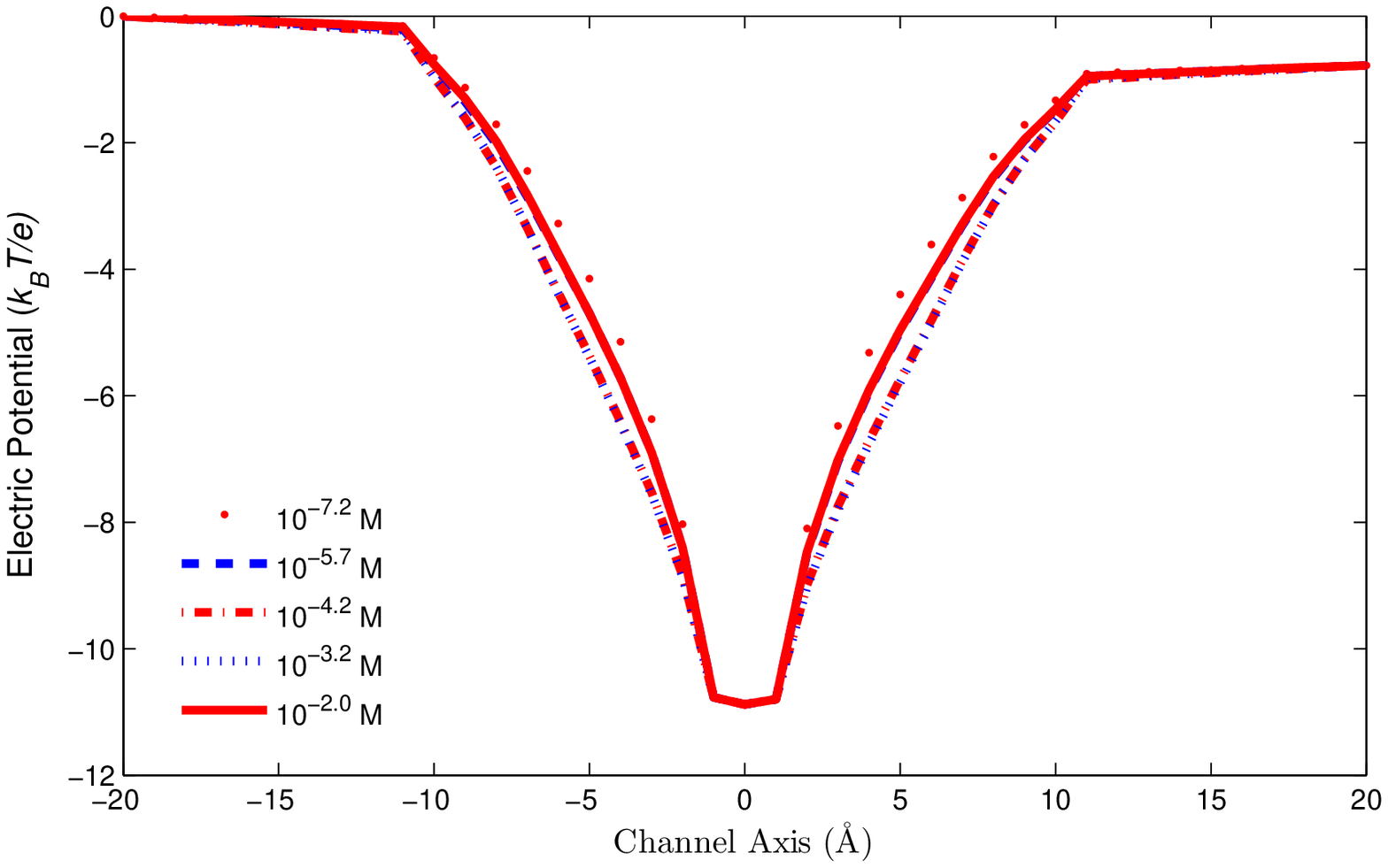}%
\caption{The averaged electric potential $\phi(\mathbf{r})$ profiles. }%
\end{figure}
\begin{figure}[ptb]%
\centering
\includegraphics[
height=2.5771in,
width=4.4114in
]%
{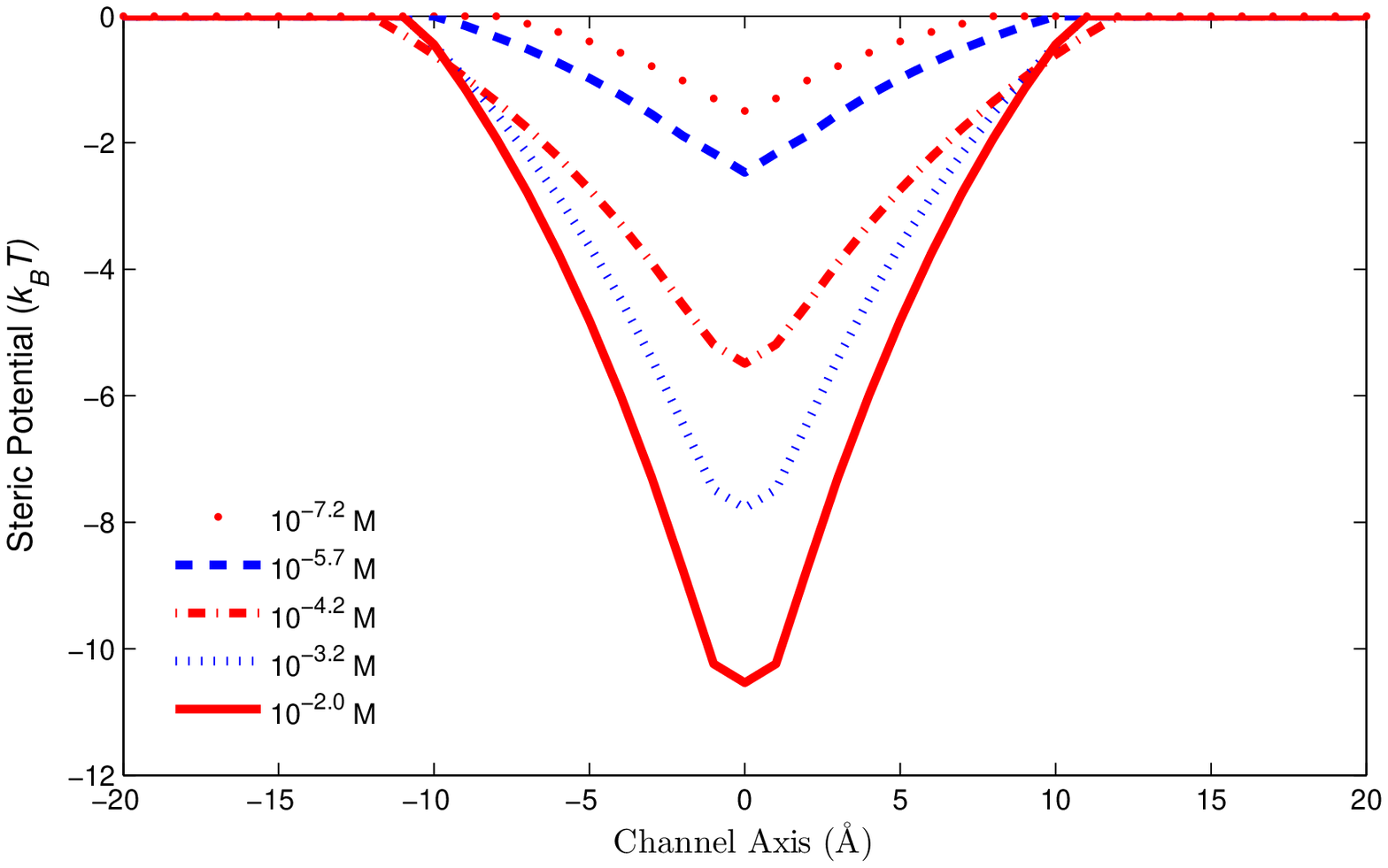}%
\caption{The averaged steric potential $S^{\text{trc}}(\mathbf{r})$ profiles.}%
\end{figure}
\begin{figure}[ptb]%
\centering
\includegraphics[
height=2.5771in,
width=4.4114in
]%
{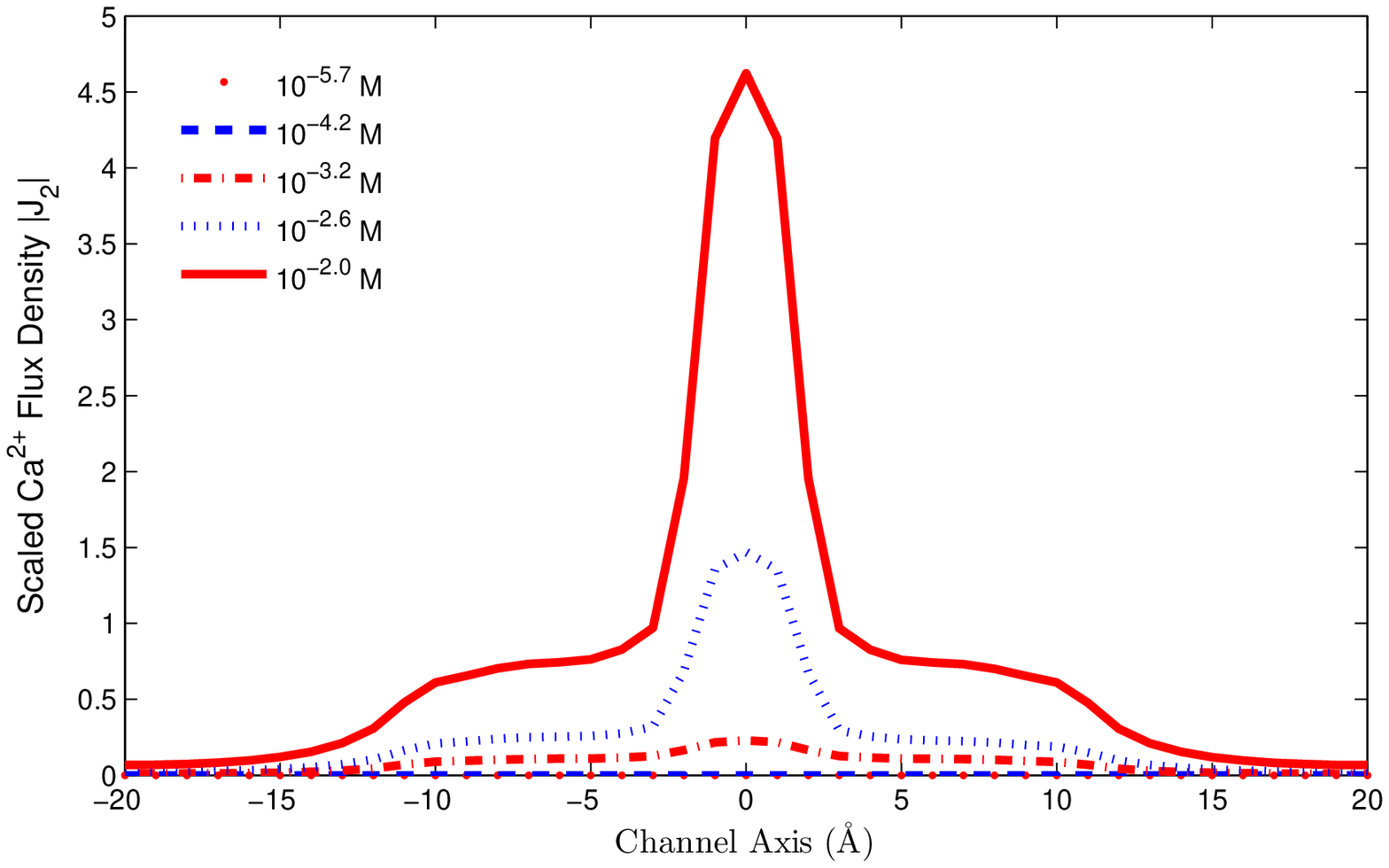}%
\caption{The averaged flux density $\left\vert \mathbf{J}_{\text{Ca}^{2+}%
}(\mathbf{r})\right\vert $ profiles.}%
\end{figure}

\textbf{\underline{Novel Features\textbf{.}}} The steric potential profiles
shown in Fig. 15 represent the novelty of the PNPF theory. All effects of
volume exclusion, interstitial void, configuration entropy, short range
interactions, correlation, polarization, screening, and dielectric response of
this nonideal system are described by the steric functional $S^{\text{trc}%
}(\mathbf{r})$. The steric potential in the binding region decreases
drastically from $-1.30$ to $-10.34$ $k_{B}T$ as [Ca$^{2+}$]$_{\text{o}}$
increases from $10^{-7.2}$ to $10^{-2}$ M. However, the electric potential
remains almost unchanged as shown in Fig. 14 following the linear occupancy
model (26). In physiological bath conditions [Ca$^{2+}$]$_{\text{o}}%
=10^{-2}\sim10^{-3}$ M, Fig. 13 shows that the region containing the binding
site with the length about 10 \AA \ is very dry (hydrophobic), which agrees
with the recent crystallographic analysis \cite{LL12} of the Ca$^{2+}$ binding
site of the related protein NCX with the EETT locus showing a hydrophobic
patch (also about 10 \AA \ in length) formed by the conserved Pro residues.
The hydrophobicity near the binding site in our model is described by the
continuous water density function $C_{\text{H}_{2}\text{O}}(\mathbf{r})$ via
the continuous steric function $S^{\text{trc}}(\mathbf{r})$, namely, the Fermi
distribution $C_{\text{H}_{2}\text{O}}(\mathbf{r})=C_{\text{H}_{2}\text{O}%
}^{\text{B}}\exp\left(  S^{\text{trc}}(\mathbf{r})\right)  $ in (5). At
[Ca$^{2+}$]$_{\text{o}}=10^{-2}$ M, the magnitude of the steric energy
$S^{\text{trc}}=-10.34$ $k_{B}T$ is comparable to that of the electrostatic
energy $\phi=-10.48$ $k_{B}T/e$. This surprisingly large energy due only to
the steric effect has not been quantified and observed by other continuum
methods in Ca$_{\text{V}}$ channel modeling, as far as we know.

The variable steric potential with respect to bath concentrations as shown in
Fig. 15 plays a similar role as the confinement potential in MC simulations
for constraining the 8 mobile O$^{1/2-}$ ions of protein glutamates in a
filter \cite{B02}. These are two different approaches to modeling the flexible
glutamates and the steric effect of ions. The excluded volumes of electrolyte
and glutamate ions are \textit{explicitly} \textit{given} as an \textit{input}
in MC simulations by using the confinement potential in a \textit{fixed}
\textit{filter} whereas the volumes are \textit{implicitly}
\textit{calculated} in PNPF simulations and are \textit{outputs} that describe
the steric potential in a \textit{variable} \textit{binding site}. We had
difficulties obtaining nonequilibrium results as those in Fig. 11 in our early
attempt to use a fixed confinement potential in a fixed filter partly because
the confinement potential would generate large artificial electric fields near
the boundary of the filter in continuum setting. It is also difficult to
incorporate the confinement potential into the flux density in Eq. (10) since
the confinement potential is fixed and cannot not be explicitly decomposed to
the electrostatic and non-electrostatic parts the way $\phi$ and
$S^{\text{trc}}$ do in (10). These difficulties are typical of inconsistent
calculations. Imposing potentials (whether in simulations or theories)
requires injection of charge and mass that is not present in the real system.
The injection occurs at a sensitive part of the system, the selectivity
filter. The approach here avoids these difficulties.

As observed from Fig. 11, ionic transport is blocked by the competition
between Na$^{+}$ and Ca$^{2+}$ ions in the range [Ca$^{2+}$]$_{\text{o}%
}=10^{-5.7}\sim10^{-4.2}$ M. In this blocking range, the corresponding steric
profiles in Fig. 15 are wider indicating that the water density or the void
volume is more evenly distributed. From Fig. 11, we observe that Ca$^{2+}$
currents increase dramatically when [Ca$^{2+}$]$_{\text{o}}$ increases from
$10^{-3.2}$ to $10^{-2}$ M in the physiological mM range of the channel, so
does the corresponding flux density as shown in Fig. 16.

\textbf{\underline{Numerical Verification\textbf{.}}} All the above results
were obtained by using the standard FD method (see \cite{L13} for instance)
for the Poisson-Fermi equations (6)-(7) and the Scharfetter-Gummel method (21)
for the flux equation (9). We now provide more numerical details for the
extended SG stability condition (18) and explain why the primitive method (12)
fails. We first verify the SG method at equilibrium ($\Delta V=V_{\text{o}%
}=V_{\text{i}}=0$) for which the PNPF solution should agree with the PF
solution as shown in Table V, where the averaged Na$^{+}$ and Ca$^{2+}$
concentrations in the filter (binding site) are denoted by $C_{1}^{\text{F}}$
and $C_{2}^{\text{F}}$, respectively. Here, the approximate PF solution of
$C_{i}(\mathbf{r})$ in (5) is treated as an exact solution to justify the
approximate PNPF solution. The PNPF concentrations agree quite well with the
PF concentrations indicating that the SG method (21) works well. Note that
$C_{1}^{\text{F}}$ and $C_{2}^{\text{F}}$ are all bounded by their respect
maximum values 462.39 and 408.57 M at very large electrostatic potential
$\phi=-10.48$ in the filter, as guaranteed by the Fermi distribution (5). We
use the SOR method \cite{GV96} for solving all linear algebraic systems with
the relaxation parameter taken to be 1.9. The error tolerance of the SOR
linear solver is 10$^{-8}$ because the boundary bath condition [Ca$^{2+}%
$]$_{\text{o}}$ for (21) is in the 10$^{8}$-fold range. The error tolerance
for solving each PDE in the PNPF model in Algorithm 2 is 10$^{-4}$.\bigskip

\begin{center}%
\begin{tabular}
[c]{l|ccccc}%
\multicolumn{6}{c}{TABLE V. Verification of the SG method (21)}\\\hline
\lbrack Ca$^{2+}$] in M & $0.9\cdot10^{-6}$ & $10^{-5}$ & $10^{-4}$ &
$10^{-3}$ & $10^{-2}$\\\hline
PF $C_{1}^{\text{F}}/C_{2}^{\text{F}}$ & 61.7/63.7 & 10.1/116.9 & 1.1/126.2 &
0.11/127.3 & 0.01/127.4\\
PNPF & 61.7/63.4 & 10.1/116.1 & 1.1/126.2 & 0.11/127.3 & 0.01/127.4\\\hline
\end{tabular}
\bigskip
\end{center}

\textbf{\underline{Primitive FD method fails\textbf{.}}} We next look more
closely into the numerics of the SG discretization concerning the SG condition
(18) at nonequilibrium ($\Delta V=-20$ mV) under the conditions as those in
Fig. 11. In Table VI, $-\beta_{i}\Delta\phi$ denotes the maximum difference of
$-\beta_{i}\phi(\mathbf{r})$ between all adjacent pairs of grid points in 3D,
where the subscript $i$ denotes the ionic species not the grid node. Since
$\left[  \text{Ca}^{2+}\right]  _{\text{o}}$ varies in the 10$^{8}$-fold
range, the maximum difference $-\beta_{i}\Delta\phi$ varies in a range for
each ionic species $i$ as shown in the table. Moreover, the adjacent pair of
grid points at which the maximum difference is obtained may differ for
different NP equations. The other two maximum differences $\Delta
S^{\text{trc}}$ and $-\beta_{i}\Delta\phi+\Delta S^{\text{trc}}$ are similar
defined. Note also that the three maximum differences may occur at different
pairs of adjacent grid points with the mesh size $h=$ 1 \AA \ even for the
same NP equation.\bigskip

\begin{center}%
\begin{tabular}
[c]{l|ccc}%
\multicolumn{4}{c}{TABLE VI. Checking the SG condition (18)}\\\hline
& $-\beta_{i}\Delta\phi$ & $\Delta S^{\text{trc}}$ & $-\beta_{i}\Delta
\phi+\Delta S^{\text{trc}}$\\\hline
NP1 (Na$^{+}$) & \multicolumn{1}{|l}{2.43 $\sim$ 2.79} &
\multicolumn{1}{l}{0.34 $\sim$ 1.75} & 1.16 $\sim$ 2.51\\
NP2 (Ca$^{2+}$) & \multicolumn{1}{|l}{4.86 $\sim$ 5.59} &
\multicolumn{1}{l}{0.34 $\sim$ 1.75} & 4.03 $\sim$ 5.30\\
NP3 (Cl$^{-}$) & \multicolumn{1}{|l}{2.47 $\sim$ 2.82} &
\multicolumn{1}{l}{0.34 $\sim$ 1.75} & 0.34 $\sim$ 1.75\\\hline
\end{tabular}
\bigskip
\end{center}

From Table VI, we observe that the primitive FD method (12) violates the SG
condition (18) for all NP1, NP2, and NP3 (without $S^{\text{trc}}$) and for
NP1 and NP2 (with $S^{\text{trc}}$). The worst case of the violation occurs in
the Ca$^{2+}$ flux equation NP2 with or without $S^{\text{trc}}$, as analyzed
in (17). Obviously, the primitive FD is not suitable for PNPF simulations on
Ca$_{\text{V}}$ channels. The SG not only delivers stable results for all
equations in the PNPF model at all experimental conditions but also converges
very rapidly in the nonlinear iteration.

\begin{center}
\textbf{VI. CONCLUSION}
\end{center}

The classical Scharfetter-Gummel (SG) method for semiconductor devices was
extended to include the steric potential for biological ion channels in this
paper. The steric potential --- a key feature of the PNPF theory ---
represents a combination of various effects of volume exclusion, interstitial
void, configuration entropy, short range interactions, correlation,
polarization, screening, and dielectric response in a complex fluid system of
ion channel. The simplified matched interface and boundary (SMIB) method
together with the SG method was shown to yield optimal convergence for a PNP
model with exact solutions of the gramicidin A channel. The primitive finite
difference method without SG was shown to lead to unphysical approximations
for an L-type calcium channel due to the violation of the generalized SG
condition presented here. Two algorithms based on the SMIB and multiscale
methods have been presented for these two different types of channels
depending on whether water is allowed to pass through the channel pore.
Numerical results for both channels are in accord with the respective
experimental results. Compared with previous PNP models, new physical details
by PNPF such as water dynamics, dielectric function, voids, and steric energy
in the system have been illustrated and discussed. The PNPF model differs from
most channel models in several respects. It computes dielectric properties as
an output that in fact vary with position and with experimental condition. It
provides a fourth order partial differential equation to describe current
flow, of the general Cahn-Hillard type, which has a richness of behavior
beyond the usual second order PNP description. Practically, it is important
that PNPF is much easier to compute in three dimensions than other steric PNP models.

\begin{center}
\textbf{ACKNOWLEDGEMENTS}
\end{center}

This work was supported in part by the Ministry of Science and Technology of
Taiwan under Grant No. 103-2115-M-134-004-MY2 to J.L.L and by the Bard Endowed
Chair of \ Rush University Medical Center, held by B.E.


\begin{thebibliography}{99}                                                                                               %
\bibitem {G64}H. K. Gummel, IEEE Trans. Elec. Dev. \textbf{11}, 163 (1964).

\bibitem {SG69}D. L. Scharfetter and H. K. Gummel, IEEE Trans. Elec. Dev.
\textbf{16}, 64 (1969).

\bibitem {S73}J. W. Slotboom, IEEE Trans. Elec. Dev. \textbf{20}, 669 (1973).

\bibitem {BR83}R. E. Bank, D. J. Rose, and W. Fichtner, IEEE Trans. Elec. Dev.
\textbf{30}, 1031 (1983).

\bibitem {MR83}P. A. Markowich, C. A. Ringhofer, S. Selberherr, and M.
Lentini, IEEE Trans. Elec. Dev. \textbf{30}, 1165 (1983).

\bibitem {S84}S. Selberherr, \textit{Analysis and Simulation of Semiconductor
Devices} (Springer-Verlag, New York, 1984).

\bibitem {S88}C. M. Snowden, \textit{Semiconductor Device Modelling} (Peter
Peregrinus Ltd., London, UK, 1988).

\bibitem {K88}T. Kerkhoven, SIAM J. Sci. Statist. Comput. \textbf{9}, 48 (1988).

\bibitem {AM89}U. Ascher, P. Markowich, C. Schmeiser, H. Steinruck, and R.
Weiss, SIAM J. Appl. Math. \textbf{49}, 165 (1989).\newline

\bibitem {BM89}F. Brezzi, L. D. Marini, and P. Pietra, SIAM J. Numer. Anal.
\textbf{26}, 1342 (1989).

\bibitem {RS89}C. Ringhofer and C. Schmeiser, SIAM J. Numer. Anal.
\textbf{26}, 507 (1989).

\bibitem {AR93}N. R. Aluru, A. Raefsky, P. M. Pinsky, K. H. Law, R. J. G.
Goossens, and R. W. Dutton, Comput. Methods Appl. Mech. Engrg. \textbf{107},
269 (1993).

\bibitem {J95}J. W. Jerome, SIAM Rev. \textbf{37}, 552 (1995).

\bibitem {KC99}M. G. Kurnikova, R. D. Coalson, P. Graf, and A. Nitzan,
Biophys. J. \textbf{76}, 642 (1999).

\bibitem {CC00}A. E. C\'{a}rdenas, R. D. Coalson, and M. G. Kurnikova,
Biophys. J. \textbf{79}, 80 (2000).

\bibitem {CL03}R.-C. Chen and J.-L. Liu, J. Comput. Phys. \textbf{189}, 579 (2003).

\bibitem {P04}R. Pinnau, SIAM J. Numer. Anal. \textbf{42}, 1648 (2004).

\bibitem {CL05}R.-C. Chen and J.-L. Liu, J. Comput. Phys. \textbf{204}, 131 (2005).

\bibitem {CL08}R.-C. Chen and J.-L. Liu, J. Comp. Phys. \textbf{227}, 6266 (2008).

\bibitem {LH10}B. Lu, M. J. Holst, J. A. McCammon, and Y. C. Zhou, J. Comput.
Phys. \textbf{229}, 6979 (2010).

\bibitem {SK10}N. A. Simakov and M. G. Kurnikova, J. Phys. Chem. B
\textbf{114}, 15180 (2010).

\bibitem {ZC11}Q. Zheng, D. Chen, and G.-W. Wei, J. Comp. Phys. \textbf{230},
5239 (2011).

\bibitem {SD99}C. Sagui and T. A. Darden, Annu. Rev. Biophys. Biomol. Struct.
\textbf{28}, 155 (1999).

\bibitem {AK00}T. W. Allen, S. Kuyucak, and S. H. Chung, Biophys. Chem.
\textbf{86}, 1 (2000).

\bibitem {CK00}B. Corry, S. Kuyucak, and S.-H. Chung, Biophys. J. \textbf{78},
2364 (2000).

\bibitem {SW01}C. N. Schutz and A. Warshel, Proteins: Structure, Function, and
Bioinformatics \textbf{44}, 400 (2001).

\bibitem {IR01}W. Im and B. Roux, Biophys. J. \textbf{115}, 4850 (2001).

\bibitem {IR02}W. Im and B. Roux, J. Mol. Biol. \textbf{322}, 851 (2002).

\bibitem {CK02}S.-H. Chung and S. Kuyucak, Biochim. Biophys. Acta
\textbf{1565}, 267 (2002).

\bibitem {EC02}S. Edwards, B. Corry, S. Kuyucak, and S.-H. Chung, Biophys. J.
\textbf{83}, 1348 (2002).

\bibitem {MJ04}G. V. Miloshevsky and P. C. Jordan, Trends in Neurosciences
\textbf{27}, 308 (2004).

\bibitem {RA04}B. Roux, T. Allen, S. Berneche, and W. Im, Q. Rev. Biophys.
\textbf{37}, 15 (2004).

\bibitem {SC11}C. Song and B. Corry, PLoSONE \textbf{6}, e21204 (2011).

\bibitem {FB02}F. Fogolari, A. Brigo, and H. Molinari, J. Mol. Recognit.
\textbf{15}, 377 (2002).

\bibitem {GK04}P. Graf, M. G. Kurnikova, R. D. Coalson, and A. Nitzan, J.
Phys. Chem. B \textbf{108}, 2006 (2004).

\bibitem {CK05}R. D. Coalson and M. G. Kurnikova, IEEE Trans. Nanobio.
\textbf{4}, 81 (2005).

\bibitem {E10}B. Eisenberg, J. Phys. Chem. C \textbf{114}, 20719 (2010).

\bibitem {E11}B. Eisenberg, in \textit{Advances in Chemical Physics}, edited
by S. A. Rice (John Wiley \& Sons, Inc. Vol. 148, 2011), p. 77.

\bibitem {E12}B. Eisenberg, SIAM News \textbf{45}, 11 (2012).

\bibitem {WZ12}G.-W. Wei, Q. Zheng, Z. Chen, and K. Xia, SIAM Rev.
\textbf{54}, 699 (2012).

\bibitem {E13}B. Eisenberg, Biophys. J. \textbf{104}, 1849 (2013).

\bibitem {LE13}J.-L. Liu and B. Eisenberg, J. Phys. Chem. B \textbf{117},
12051 (2013).

\bibitem {LE14b}J.-L. Liu and B. Eisenberg, J. Chem. Phys. \textbf{141},
22D532 (2014).

\bibitem {GY07}W. Geng, S. Yu, and G. Wei, J. Chem. Phys. \textbf{127}, 114106 (2007).

\bibitem {L13}J.-L. Liu, J. Comp. Phys. \textbf{247}, 88 (2013).

\bibitem {H01}B. Hille, \textit{Ionic Channels of Excitable Membranes}
(Sinauer Associates Inc., Sunderland, MA, 2001).

\bibitem {M39}N. F. Mott, The theory of crystal rectifiers, Proc. Roy. Soc. A
\textbf{171}, 27 (1939).

\bibitem {SM03}W. A. Sather and E. W. McCleskey, Annu. Rev. Physiol.
\textbf{65}, 133 (2003).

\bibitem {S06}C. D. Santangelo, Phys. Rev. E \textbf{73}, 041512 (2006).

\bibitem {BS11}M. Z. Bazant, B. D. Storey, and A. A. Kornyshev, Phys. Rev.
Lett. \textbf{106}, 046102 (2011).

\bibitem {HL05}S. M. Hou and X.-D. Liu, J. Comput. Phys. \textbf{202}, 411 (2005).

\bibitem {CB92}D. P. Chen, V. Barcilon, and R. S. Eisenberg, Biophys. J.
\textbf{61}, 1372 (1992).

\bibitem {EC93}R. Eisenberg and D. Chen, Biophys. J. \textbf{64}, A22 (1993).

\bibitem {EK95}R. S. Eisenberg, M. M. K\l osek, and Z. Schuss, J. Chem. Phys.
\textbf{102}, 1767 (1995).

\bibitem {FZ06}S. Furini, F. Zerbetto, and S. Cavalcanti, Biophys. J.
\textbf{91}, 3162 (2006).

\bibitem {G08}D. Gillespie, Biophys. J. \textbf{94}, 1169 (2008).

\bibitem {FL63}R. P. Feynman, R. B. Leighton, and M. Sands, \textit{The
Feynman Lectures on Physics, Volume II, Mainly Electromagnetism and Matter}
(Addison-Wesley Publishing Co., New York, 1963).

\bibitem {BV09}D. Boda, M. Valisk\'{o}, D. Henderson, B. Eisenberg, D.
Gillespie, and W. Nonner, J. Gen. Physiol. \textbf{133}, 497 (2009).

\bibitem {B02}H. M. Berman et al., Acta Cryst. \textbf{D58}, 899 (2002).

\bibitem {SR73}A. Shrake and J. A. Rupley, J. Mol. Biol. \textbf{79}, 351 (1973).

\bibitem {NH03}B. Nadler, U. Hollerbach, and R. S. Eisenberg, Phys. Rev. E
\textbf{68} (2003) 021905.

\bibitem {BB02}D. Boda, D. D. Busath, B. Eisenberg, D. Henderson, and W.
Nonner, Phys. Chem. Chem. Phys. \textbf{4}, 5154 (2002).

\bibitem {BN07}D. Boda, W. Nonner, M. Valisk\'{o}, D. Henderson, B. Eisenberg,
and D. Gillespie, Biophys. J. \textbf{93}, 1960 (2007).

\bibitem {LF01}G. M. Lipkind and H. A. Fozzard, Biochem. \textbf{40}, 6786 (2001).

\bibitem {LE14}J.-L. Liu and B. Eisenberg, J. Chem. Phys. \textbf{141}, 075102 (2014).

\bibitem {AM84}W. Almers and E. W. McCleskey, J. Physiol. \textbf{353}, 585 (1984).

\bibitem {LL12}J. Liao et al., Science \textbf{335}, 686 (2012).

\bibitem {DM98}D. A. Doyle et al., Science \textbf{280}, 69 (1998).

\bibitem {LC13}M. Liao, E. Cao, D. Julius, and Y. Cheng, Nature \textbf{504},
107 (2013).

\bibitem {TC14}L. Tang, L. Tang, T. M. G. El-Din, J. Payandeh, G. Q. Martinez,
T. M. Heard, T. Scheuer, N. Zheng, and W. A. Catterall, Nature \textbf{505},
56 (2014).

\bibitem {DC07}T. J. Dolinsky, P. Czodrowski, H. Li, J. E. Nielsen, J. H.
Jensen, G. Klebe, and N. A. Baker, Nucleic Acids Res. \textbf{35}, W522 (2007).

\bibitem {HC01}U. Hollerbach, D.-P. Chen, and R. S. Eisenberg, J. Scient.
Comput. \textbf{16}, 373 (2001).

\bibitem {CF02}C. D. Cole, A. S. Frost, N. Thompson, M. Cotten, T. A. Cross,
and D. D. Busath, Biophys. J. \textbf{83}, 1974 (2002).

\bibitem {MC03}A. Mamonov, R. D. Coalson, A. Nitzan, and M. G. Kurnikova,
Biophys. J. \textbf{84}, 3646 (2003).

\bibitem {SS98}G. R. Smith and M. S. P. Sansom, Biophys. J. \textbf{75}, 2767 (1998).

\bibitem {LM08}B. Lu and J. A. McCammon, Chem Phys Lett. \textbf{451}, 282 (2008).

\bibitem {RP95}B. Roux, B. Prod'hom, and M. Karplus, Biophys. J. \textbf{68},
876 (1995).

\bibitem {R02}B. Roux, Acc. Chem. Res. \textbf{35}, 366 (2002).

\bibitem {BK06}T. Ba\c{s}tu\u{g} and S. Kuyucak, Biophys. J. \textbf{90}, 3941 (2006).

\bibitem {BK09}D. Bucher and S. Kuyucak, Chem. Phys. Lett. \textbf{477}, 207 (2009).

\bibitem {GV96}G. Golub and C. van Loan, \textit{Matrix Computations} (The
Johns Hopkins University Press 1996).
\end{thebibliography}
\end{document}